\documentclass[a4paper,11pt]{article}
\usepackage{jheppub}
\usepackage{amsthm}
\usepackage{tikz}
\usetikzlibrary{fit}
\usepackage{braket}
\usepackage{cancel}

\newcommand{\bZ}{\mathbb{Z}}
\newcommand{\bR}{\mathbb{R}}

\usepackage{pict2e}
\makeatletter
\newcommand{\farsquare}[2]{#1\,{\mathpalette\far@square{#2}}}
\newcommand{\far@square}[2]{%
  \mathop{\vcenter{\hbox{%
    \sbox\z@{$\m@th#1\sum$}%
    \setlength{\unitlength}{0.9\dimexpr\ht\z@+\dp\z@}%
    \begin{picture}(1,1)
    \roundjoin
    \polyline(0,0)(0,1)(1,1)(1,0)(0,0)(0,0.5)
    \end{picture}%
  }}}\limits_{#1#2}%
}
\makeatother

\makeatletter

\newcommand{\cubeBC}[3]{\mathpalette\cubeBC@{{#1}{#2}{#3}}}

\newcommand{\cubeBC@}[2]{%
\mathord{\vcenter{\hbox{%
\setlength{\unitlength}{2.6ex}%
\begin{picture}(2.4,2.4)
\roundjoin

\polyline(0.35,0.35)(0.35,1.35)(1.35,1.35)(1.35,0.35)(0.35,0.35)
\polyline(0.80,0.80)(0.80,1.80)(1.80,1.80)(1.80,0.80)(0.80,0.80)
\polyline(0.35,0.35)(0.80,0.80)
\polyline(0.35,1.35)(0.80,1.80)
\polyline(1.35,1.35)(1.80,1.80)
\polyline(1.35,0.35)(1.80,0.80)

\put(0.18,0.85){\makebox(0,0)[r]{$\scriptstyle\firstofthree#2$}} 
\put(0.85,0.18){\makebox(0,0)[t]{$\scriptstyle\secondofthree#2$}} 
\put(1.65,0.35){\makebox(0,0)[l]{$\scriptstyle\thirdofthree#2$}}  

\end{picture}
}}}
}

\newcommand{\firstofthree}[3]{#1}
\newcommand{\secondofthree}[3]{#2}
\newcommand{\thirdofthree}[3]{#3}

\makeatother

\usepackage{pgfplots}
\pgfplotsset{compat=1.18}
\usepackage{tikz}

\makeatletter
\def\@fpheader{}
\makeatother

\title{M-theory on $S^1\vee S^1$ as Type 0A}

\author[1]{Zihni Kaan Baykara,}
\author[2]{Emilian Dudas,}
\author[1]{Cumrun Vafa}
\affiliation[1]{Jefferson Physical Laboratory, Harvard University,\\
Cambridge, MA 02138, USA}
\affiliation[2]{CPHT, CNRS,
Ecole polytechnique, Institut Polytechnique de Paris,\\
91120 Palaiseau, FRANCE}

\emailAdd{zbaykara@g.harvard.edu, emilian.dudas@polytechnique.edu, vafa@g.harvard.edu}

\abstract{We propose an exotic geometric M-theory dual for the weak coupling Type 0A string: compactification on a sub-Planckian $S^1\vee S^1$ (two circles connected at a point), where strong quantum effects lead to fields living on distinct resolutions of that space.  Moreover we argue that tachyon condensation of the 0A theory corresponds to shrinking of one of the two circles leading to the IIA supersymmetric string. We use this and other dualities to provide an F-theoretic description of the axio-dilaton and the tachyonic field of Type 0B and argue for the existence of a strong coupling critical point of the potential using the resulting duality symmetry $\Gamma_0(2)\subset SL(2,\bZ)$ when tachyon field vanishes. The existence of this critical point can also be argued using conventional M-theory dualities.  If this critical point is unique it is an unstable dS vacuum.  Using this we propose a strong coupling conformal fixed point for a non-supersymmetric gauge theory in four dimensions living on coincident $D3^+-D3^-$branes of 0B.  More generally we conjecture that the tachyon field can be identified with a pair of points on the torus parameterized by $z$ leading to the full duality group $SL(2,\bZ)$ acting on $(z,\tau)$.  We find tantalizing hints that for both 0A and 0B the tachyon fields can be identified with the holonomy of a flat $SU(2)$ gauge field suggesting that we have two copies of the universe interacting with one another as if they are D-branes!}
\begin{document}
\maketitle
\flushbottom

\section{Introduction}
We live in a universe with positive vacuum energy with broken supersymmetry.  Despite progress on multiple fronts in understanding string theory over many decades, it is fair to say that our understanding of non-supersymmetric backgrounds in string theory and their strong coupling dynamics as well as instabilities is very limited.\footnote{For some reviews, see e.g. \cite{leone-raucci}, \cite{augusto-rev}, \cite{carlo}.}  In particular, despite many attempts to find (meta-)stable non-supersymmetric vacua, no viable candidate has been identified.  Moreover, related ideas such as the Swampland dS conjecture \cite{Obied:2018sgi} and the TCC \cite{Bedroya:2019snp} (see also \cite{Bedroya:2022tbh,Bedroya:2024zta}) have led to an explanation as to why it may be fundamentally impossible to find one.  See \cite{Vafa:2025nst,Agmon:2022thq,Palti:2019pca,vanBeest:2021lhn,Grana:2021zvf,Lehnert:2025izp} for reviews of some of these ideas.  Moreover, stringy ideas and instabilities of non-supersymmetric vacua lead to the natural expectation that dark matter and dark energy evolve in time. This has led to a simple predictive model \cite{Agrawal:2019dlm,Bedroya:2025fwh} which has a good fit with the recent astrophysical observations \cite{DES:2024jxu, DES:2025tna,DESI:2025fii, DESI:2025zgx, DESI:2025zpo}. Therefore attempts to find metastable non-supersymmetric quantum gravity vacua may be neither possible, nor desirable!

Given this state of affairs, it is important to better understand non-supersymmetric string vacua and the fate of their instabilities and tachyon condensations leading to evolution towards stable vacua.
Perhaps the simplest examples of non-supersymmetric superstring vacua are Type 0A and Type 0B string theories in 10 dimensions.  These theories have no light fermions in their spectra and in addition have tachyonic modes.  Our main aim is to demystify strong coupling features of both of these theories as well as propose a picture for their tachyon condensation.

Here we propose and find tantalizing evidence that the M-theory lift of weak coupling Type 0A theory is on a singular 1-dimensional sub-Planckian space:  $S^1\vee S^1$, i.e. the wedge sum of two circles touching at a point as in figure `8', with the points where circles join being a frozen singularity. We propose what this singular geometry means and what its quantum realization may mean. This picture geometrizes the field content of 0A theory, including the tachyonic mode which controls the relative perimeters of the two circles.  Tachyon condensation corresponds to shrinking away one of the two circles, leading to Type IIA strings.  We present several pieces of evidence for this picture many of which come from the study of D-branes in 0A.  We also find hints why making the circles macroscopically large is obstructed by a potential, consistent with quantum corrections playing a crucial role in making sense of this space.

We then apply this picture to find a geometric F-theoretic description of Type 0B. In particular we find that when tachyon field vanishes the theory has $\Gamma_0(2)$ (the level 2 subgroup of $SL(2,\bZ)$) S-duality symmetry.  We argue for this S-duality also using more conventional M-theory dualities.  We also find a critical point of the theory at a strong coupling point $\tau=\frac{i+1}{2}$ and argue it is an unstable dS vacuum if it is unique.  We propose a non-supersymmetric conformal 4d theory of pairs of $D3^\pm$branes of this theory at this value of the gauge coupling. More generally we conjecture that the tachyon field can be viewed as a pair of points on the torus parameterized by $z$ which splits the coupling between the two gauge factors leading to $SL(2,\bZ)$ as the full duality group of type 0B theory.  In addition we argue that the tachyon condensation of this theory (taking $z\rightarrow 0$) leads to Type IIB.  We find tantalizing hints that for both 0A and 0B the tachyon fields can be identified with the holonomy of a flat $SU(2)$ gauge field suggesting that we have two copies of the universe interacting with one another as if they are D-branes!

The organization of this paper is as follows:
In \autoref{sec:Type0} we review Type 0A and 0B theories and their connections. In \autoref{sec:0A-arguments} we show why some previous proposals for connecting 0A with M-theory do not work.
In \autoref{sec:S1vS1} we motivate our conjecture for the M-theory description of 0A, provide evidence for this picture and explain the fate of tachyon condensation for 0A. In \autoref{sec:F-0B} we discuss connections with 0B theory, implications for its strong coupling points and tachyon condensation.  
In \autoref{sec:conclusion} we end with some concluding thoughts. Some technical aspects of freely acting orbifolds that we use in the paper are presented in \autoref{app:orbifold}.

\section{Type 0 strings}\label{sec:Type0}
\subsection{Spectrum}\label{sec:spectrum}

Type 0A and 0B string theories \cite{Dixon:1986iz} are obtained from the RNS formulation by modifying the GSO projection of Type II strings, or equivalently by orbifolding IIA and IIB by $(-1)^{F}$ where $F$ is the spacetime fermion number. This leads to a theory in which all fermions are removed and the RR fields are doubled.  In addition, both theories have an extra bosonic field which is tachyonic.

Let us describe the spectrum in more detail using the RNS formalism.\footnote{In the Green-Schwarz formalism this corresponds to having both chiralities of spinors for left and right-movers in the RR sectors.  For Type 0A the left and right spinors will have opposite chirality whereas in 0B they will have the same chirality.} Instead of imposing a chiral GSO projection, one considers identical spin structures for fermions in the left and right sectors. To accomplish this from Type II perspective one projects onto states invariant under $(-1)^{G_L+G_R}$ in the NSNS sector and $\pm 1$ in the RR sector, where $G_L$ and $G_R$ are left and right worldsheet fermion numbers. As usual, the superscripts $+$ and $-$ denote the two signs of GSO projections $P_{GSO} = \frac{1\pm (-1)^{G_{L,R}}}{2}$ in a given sector.

Under this projection the mixed NSR and RNS sectors are removed, and therefore the perturbative spectrum contains no spacetime fermions. The surviving NSNS sectors are $({\rm NS}^+,{\rm NS}^+)$ and $({\rm NS}^-,{\rm NS}^-)$, where the latter contains a tachyon. In the RR sector, Type 0B contains $({\rm R}^+,{\rm R}^+)$ and $({\rm R}^-,{\rm R}^-)$, while Type 0A contains $({\rm R}^+,{\rm R}^-)$ and $({\rm R}^-,{\rm R}^+)$. Unlike in Type II, where the GSO projection selects a single Ramond sector, the diagonal projection used in Type 0 retains both allowed Ramond sectors. The RR spectrum is therefore doubled, and we denote the second copy of RR gauge potentials by primed fields. The field content of the Type 0A and Type 0B theories is summarized in \autoref{fig:type0spectra}.

\begin{table}[t]
\centering
\begin{tabular}{|c|c|c|}
\hline
Sector & Type 0B & Type 0A \\
\hline
$({\rm NS}^+,{\rm NS}^+)$ 
& $g_{\mu\nu},\; B_{\mu\nu},\; \phi$ 
& $g_{\mu\nu},\; B_{\mu\nu},\; \phi$ \\

$({\rm NS}^-,{\rm NS}^-)$ 
& $T$ 
& $T$ \\

$({\rm R}^+,{\rm R}^+)$ 
& $\chi,\; C_{\mu\nu},\; D_{\mu\nu\rho\sigma}$ 
& --- \\

$({\rm R}^-,{\rm R}^-)$ 
& $\chi',\; C_{\mu\nu}',\; D_{\mu\nu\rho\sigma}'$ 
& --- \\

$({\rm R}^+,{\rm R}^-)$ 
& --- 
& $A_\mu,\; C_{\mu\nu\rho}$ \\

$({\rm R}^-,{\rm R}^+)$ 
& --- 
& $A_\mu',\; C_{\mu\nu\rho}'$ \\
\hline
\end{tabular}
\caption{Perturbative spectra of Type 0B and Type 0A string theories. The NSNS sector contains $g_{\mu\nu}, B_{\mu\nu}, \phi$, and a tachyon $T$. The RR sector is doubled relative to Type II, giving two RR gauge fields.}
\label{fig:type0spectra}
\end{table}

Type 0 strings can also be understood as orbifolds of Type II theories by the spacetime fermion number operator $(-1)^F$. The orbifold description implies the existence of a $\mathbb{Z}_2$ quantum symmetry $Q$ acting on the twisted sector states, implemented by $Q=(-1)^{G_L}$.\footnote{Orbifolding further by the quantum symmetry recovers the parent theory.} This symmetry acts trivially in the $(NS^+,NS^+)$ sector and in the RR sector common with Type II, and by $(-1)$ on the $(NS^-,NS^-)$ and on the second RR sector as $C_p'\to -C_p'$. It is convenient to form linear combinations of the $p$-form gauge fields as
\begin{align}\label{eq:Cppm}
    C_p^\pm \equiv {C_p\pm C_p' \over {2}},
\end{align}
so that the quantum symmetry acts as an exchange $C_p^+\leftrightarrow C_p^-$. Overall, the quantum symmetry acts on the massless and tachyonic fields as
\begin{align}
Q:\quad
T &\to -T, \qquad
A^+ \leftrightarrow A^-, \qquad
C_3^+ \leftrightarrow C_3^-
&& ({\rm 0A}) \\
Q:\quad
T &\to -T, \qquad\chi^+\leftrightarrow \chi^-,\qquad 
C_2^+ \leftrightarrow C_2^-, \qquad
D_4^+ \leftrightarrow D_4^-
&& ({\rm 0B})
\end{align}
and trivially on the omitted massless fields.

\begin{figure}[t]
\centering
\begin{tikzpicture}[>=stealth]

\node (iia_parent_title) at (0,6.0) {{\bf IIA}};

\node (iia_p_nsns) at (0,5.3) {$({\rm NS}^+,{\rm NS}^+)$};
\node (iia_p_nsrs) at (0,4.8) {$\cancel{({\rm NS}^+,{\rm R}^-)}$};
\node (iia_p_rns)  at (0,4.3) {$\cancel{({\rm R}^+,{\rm NS}^+)}$};
\node (iia_p_rr)   at (0,3.8) {$({\rm R}^+,{\rm R}^-)$};

\node (iiap_parent_title) at (8,6.0) {$\mathbf{IIA'}$};

\node (iiap_p_nsns) at (8,5.3) {$({\rm NS}^+,{\rm NS}^+)$};
\node (iiap_p_nsrs) at (8,4.8) {$\cancel{({\rm NS}^+,{\rm R}^+)}$};
\node (iiap_p_rns)  at (8,4.3) {$\cancel{({\rm R}^-,{\rm NS}^+)}$};
\node (iiap_p_rr)   at (8,3.8) {$({\rm R}^-,{\rm R}^+)$};

\draw[<->,dashed,bend left=35] (iia_parent_title) to (iiap_parent_title);
\node at (4,6.8) {$\mathbb Z_2$};

\draw[->,dashed] (iia_p_rr) -- (0,2.5);
\draw[->,dashed] (iiap_p_rr) -- (8,2.5);

\node (iia_title) at (0,2.2) {$\mathbf{IIA/(-1)^F}$};

\node (iia_ns_u) at (0,1.3) {$({\rm NS}^+,{\rm NS}^+)$};
\node (iia_rr_u) at (0,0.8) {$({\rm R}^+,{\rm R}^-)$};

\node (iia_ns_t) at (0,-0.7) {$({\rm NS}^-,{\rm NS}^-)$};
\node (iia_rr_t) at (0,-0.2) {$({\rm R}^-,{\rm R}^+)$};

\node[draw,rounded corners,fit=(iia_ns_u)(iia_rr_u),label=left:{untwisted}] {};
\node[draw,rounded corners,fit=(iia_ns_t)(iia_rr_t),label=left:{twisted}] {};

\node (iiap_title) at (8,2.2) {$\mathbf{IIA'/(-1)^F}$};

\node (iiap_ns_u) at (8,1.3) {$({\rm NS}^+,{\rm NS}^+)$};
\node (iiap_rr_u) at (8,0.8) {$({\rm R}^-,{\rm R}^+)$};

\node (iiap_ns_t) at (8,-0.7) {$({\rm NS}^-,{\rm NS}^-)$};
\node (iiap_rr_t) at (8,-0.2) {$({\rm R}^+,{\rm R}^-)$};

\node[draw,rounded corners,fit=(iiap_ns_u)(iiap_rr_u),label=right:{untwisted}] {};
\node[draw,rounded corners,fit=(iiap_ns_t)(iiap_rr_t),label=right:{twisted}] {};

\node (zero_title) at (4,2.2) {{\bf Type 0A}};

\node (z_ns_u) at (4,1.3) {$({\rm NS}^+,{\rm NS}^+)$};
\node (z_rr_1) at (4,0.8) {$({\rm R}^+,{\rm R}^-)$};
\node (z_ns_t) at (4,-0.7) {$({\rm NS}^-,{\rm NS}^-)$};
\node (z_rr_2) at (4,-0.2) {$({\rm R}^-,{\rm R}^+)$};

\draw[->] (iia_ns_u) -- (z_ns_u);
\draw[->] (iia_rr_u) -- (z_rr_1);
\draw[->] (iia_ns_t) -- (z_ns_t);
\draw[->] (iia_rr_t) -- (z_rr_2);

\draw[->] (iiap_ns_u) -- (z_ns_u);
\draw[->] (iiap_ns_t) -- (z_ns_t);
\draw[->] (iiap_rr_u.west) -- (z_rr_2.east);
\draw[->] (iiap_rr_t.west) -- (z_rr_1.east);

\draw[<->,dashed] (zero_title) to[out=30,in=150,looseness=8] (zero_title);
\node at (4,3.3) {$\mathbb Z_2$};

\node at (2,2.2) {$=$};
\node at (6,2.2) {$=$};

\end{tikzpicture}
\caption{Type 0A obtained as the orbifold of Type IIA or Type ${\rm IIA}'$ by $(-1)^F$. The theory ${\rm IIA}'$ is related to IIA by the left–right mover exchange $\Omega$. This becomes a $\mathbb{Z}_2$ symmetry of Type 0A. The $(-1)^F$ orbifolds differ by an exchange of twisted and untwisted RR sectors.}
\label{fig:type0a-orb}
\end{figure}
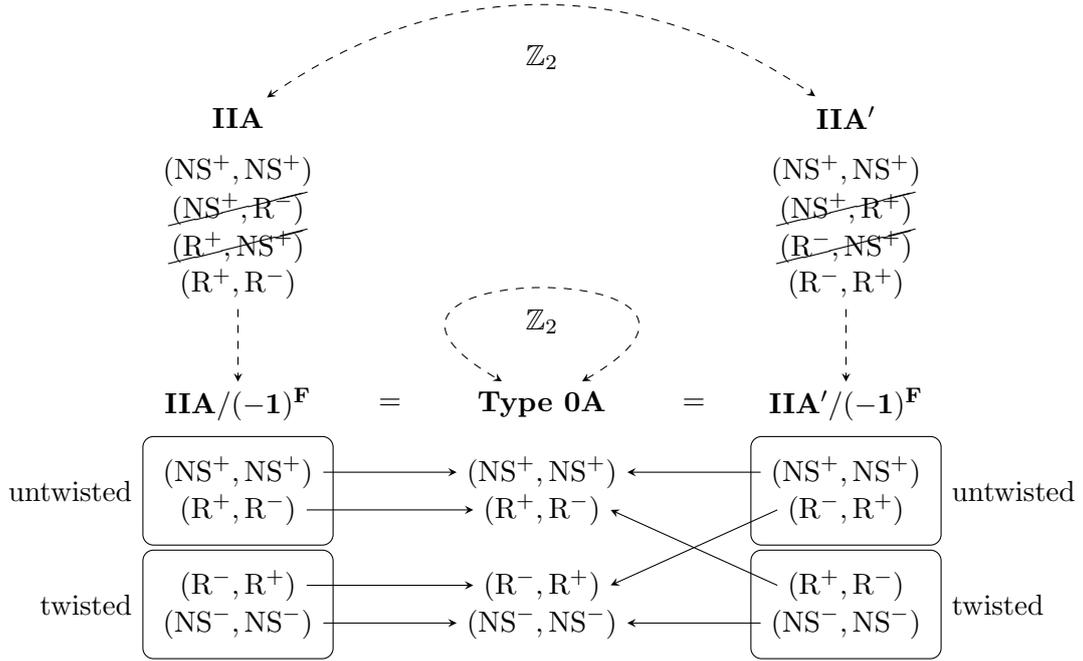

The orbifold perspective also clarifies the origin of the two RR sectors in Type 0A. In addition to the usual Type IIA theory, one may consider Type ${\rm IIA}'$ theory obtained by acting with the worldsheet parity operator $\Omega$ on Type IIA, which flips the chiralities of the left- and right-moving Ramond sectors. Orbifolding either IIA or ${\rm IIA}'$ by $(-1)^F$ produces the same Type 0A theory. As a result, $\Omega$ is promoted to a $\mathbb Z_2$ symmetry in Type 0A,
\begin{align}
\Omega:
A^\pm \to \pm A^\pm, \qquad
C_3^\pm \to \pm C_3^\pm,\qquad B_2 \to -B_2 && ({\rm 0A})
\end{align}
with trivial action on other massless fields and the tachyon, whereas $\Omega$ used to only be a map exchanging the parent theories $\Omega:{\rm IIA}\leftrightarrow {\rm IIA}'$. So 0A may be viewed as sitting “in between” IIA and ${\rm IIA}'$. The two constructions differ only in the origin of the RR states: the untwisted RR sector of the IIA orbifold corresponds to the twisted RR sector of the ${\rm IIA’}$ orbifold, and vice versa. These relations between IIA, ${\rm IIA}'$, and Type 0A are summarized in \autoref{fig:type0a-orb}.

\subsection{D-branes}\label{sec:branes}
\subsubsection{Spectrum}
The doubling of the RR sector implies that the D-brane spectrum of Type 0A and Type 0B is likewise doubled relative to the corresponding Type II theories. Similar to \eqref{eq:Cppm}, the charges are redefined from $q,q'$ associated with $C_p,C_p'$ to the linear combination
\begin{align}
    q_\pm \equiv {q\pm q' \over 2}.
\end{align}

Boundary states provide a convenient worldsheet description of D-branes. In this formalism a D$p$-brane is represented by a state in the closed-string Hilbert space which implements the appropriate boundary conditions for open strings ending on the brane. See \cite{Bergman:1997rf, Klebanov:1998yya, Gaberdiel:2000jr, Thompson:2001rw} for the boundary states of Type 0.

The boundary state decomposes into contributions from the different closed-string sectors, labelled by the NSNS or RR sectors and by a choice of $\eta=\pm$ satisfying
\begin{align}
    (\psi^\mu_L - i \eta S^\mu{}_\nu \psi_R^\nu)\ket{Bp,\eta}_{\mathrm{NSNS},\mathrm{RR}}=0,
\end{align}
where $S^{\mu\nu}= (\eta^{\alpha\beta},-\delta^{ij})$ with $\alpha,\beta$ in the worldvolume and $i,j$ in the transverse directions.

\begin{table}[t]
\centering
\begin{tabular}{|c|c|c|}
\hline
Endpoints & Sector & Fields \\
\hline
$Dp^\pm - Dp^\pm$ & NS$^+$ & $A_\mu,\ \phi^i$ \\
$Dp^\pm - \overline{Dp^\pm}$ & NS$^-$ & $T$ \\
$Dp^+ - Dp^-$ & R$^+$ & $\psi$ \\
$Dp^+ - \overline{Dp^-}$ & R$^-$ & $\psi^c$ \\
\hline
\end{tabular}
\caption{Open string spectrum between Type 0 D-branes. NS$^+$ contains the massless bosons $A_\mu$ and scalars $\phi^i$ for $i=1,\dots,9-p$. NS$^-$ contains the tachyon $T$. The Ramond sectors contain spacetime fermions where $R^+$ and $R^-$ correspond to spinors and conjugate spinors $\psi$ and $\psi^c$.}
\label{fig:open-spectrum}
\end{table}

After GSO projection and imposing consistency conditions, it turns out that in Type 0 we get
\begin{align}
\begin{split}
    \ket{Dp^+}&= \ket{Bp,+}_{\rm NSNS} + \ket{Bp,+}_{\rm RR},\\
    \ket{Dp^-}&= \ket{Bp,-}_{\rm NSNS} + \ket{Bp,-}_{\rm RR},\\
    \ket{\overline{Dp^+}}&= \ket{Bp,+}_{\rm NSNS} - \ket{Bp,+}_{\rm RR},\\
    \ket{\overline{Dp^-}}&= \ket{Bp,-}_{\rm NSNS} - \ket{Bp,-}_{\rm RR},\\
\end{split}
\end{align}
where $p$ is even in Type 0A and odd in Type 0B.\footnote{Also there are unstable neutral branes $\ket{\widehat{Dp^+}}=\ket{Bp,+}_{\rm NSNS}$ and $\ket{\widehat{Dp^-}}=\ket{Bp,-}_{\rm NSNS}$ for $p$ odd in Type 0A and $p$ even in Type 0B. These $\widehat{Dp^\pm}$ branes are unstable not only in the BPS sense but in the sense that they have tachyonic open strings living on them, whereas $Dp^\pm$ do not \cite{dms}.} Thus for each allowed value of $p$, there exist two distinct D-branes $Dp^+, Dp^-$ together with the anti-branes $\overline{Dp^+},\overline{Dp^-}$, coupling to $C_{p+1}^\pm$ respectively.
\medskip

The open string spectra follow from annulus amplitudes between D-branes, and transform under the worldvolume gauge groups. The open string spectrum in $Dp^+-Dp^+$ and $Dp^--Dp^-$ is purely bosonic, consisting of a gauge field and transverse scalars. Between $Dp^+-\overline{Dp^+}$ and similarly $Dp^--\overline{Dp^-}$ we have tachyons corresponding to the annihilation mode of the branes. Between unlike branes $Dp^+-Dp^-$ we get spinors $\psi$ and for $Dp^+-\overline{Dp^-}$ we get conjugate spinors $\psi^c$. The spectrum is summarized in \autoref{fig:open-spectrum}.

\subsubsection{Effective tension}
It was shown in \cite{Klebanov:1998yya} that the tension of the stable Type 0 D-branes is given by that of Type II branes divided by $\sqrt{2}$, 
\begin{align}
    T_{Dp^\pm}^{(0)} = \frac{1}{\sqrt{2}(2\pi)^{p}}\frac{M_s^{p+1}}{\lambda_s},
\end{align}
for even $p$ in Type 0A and odd $p$ in Type 0B, where $\lambda_s$ is the string coupling and $M_s$ is the string mass.

The difference between the two types of branes arises in their couplings to the NSNS sector. In particular, the boundary states $|Bp,+\rangle_{\text{NSNS}},|Bp,-\rangle_{\text{NSNS}}
$ couple with opposite sign to the twisted NSNS sector, which contains the tachyon.
At the level of the worldvolume effective action we have \cite{Klebanov:1998yya, Garousi:1999fu} 
\begin{align}
S_{Dp^\pm}=
-T_p^{(0)}\int d^{p+1}x \sqrt{-\hat G}\;e^{-\phi}\left(1 \mp \frac{1}{4} T +\frac{3}{32} T^2+\dots\right),
\end{align}
where $T$ denotes the canonically normalized tachyon and $\hat G$ the induced metric.\footnote{Our tachyon $T$ is the negative of the convention used in \cite{Klebanov:1998yya}.}
Thus the two types of branes have identical bare tensions and equal magnitude RR charge, but opposite linear couplings to the tachyon. A non-vanishing tachyon expectation value therefore shifts their effective tensions oppositely,
\begin{align}
    T_{Dp^\pm}^{\text{eff}}=
\frac{1}{\sqrt{2}(2\pi)^p}\frac{M_s^{p+1}}{\lambda_s}\left(1 \mp \frac 1 4 T + \frac{3}{32}T^2+\dots\right).
\end{align}
Garousi has conjectured in \cite{Garousi:1999fu} that the full expression is
\begin{align}
    T^{\rm eff}_{Dp^\pm} = \frac{1}{\sqrt{2}(2\pi)^p} \frac{M_s^{p+1}}{\lambda_s \sqrt{1\pm \frac T 2}}.
\end{align}
The relative sign will play an important role later in our discussion of tachyon condensation.

\subsection{T-duality}\label{sec:T-duality}
In this section we apply the results on T-duality of freely-acting orbifolds derived in \autoref{app:orbifold} to explain the perturbative dualities between Type 0 and II strings. 

Consider a symmetry $g$ of a Type II or 0 string theory. One may compactify on a circle with $g$-twisted boundary conditions, i.e. on a ``twisted circle'' $S^1_g$, where a shift around the circle is accompanied by the action of $g$. Equivalently, this is the freely-acting orbifold
\begin{align}
S^1_g \equiv \frac{ S^1}{g \cdot T_\delta},
\end{align}
where $\delta$ is a shift along the circle, and $T_\delta$ is the corresponding operator on the Hilbert space.
T-duality maps a Type A and Type B theory on such a twisted circle as
\begin{align}\label{eq:t-duality}
\text{Type A on } S^1_g \Big|_{R}
=\frac{\text{Type B}}{g}\text{ on } S^1_Q \Big|_{\frac{N}{R}},
\end{align}
where $N$ is the order of $g$ and $Q$ is the quantum symmetry of the orbifolded theory $\text{Type B}/g$, which acts by a phase $e^{2\pi i k/N}$ in the $g^k$-twisted sector.

As a trivial example, take $g=1$. Then $S^1_g$ is an ordinary circle, and T-duality reduces to the familiar relation
\begin{align}\label{eq:0a-0b}
\text{0A on } S^1\Big|_{R}=
\text{0B on } S^1\Big|_{\frac{1}{R}},
\end{align}
which follows directly from the reversal of the right-moving Ramond chirality and exchange of momentum-winding under T-duality.

A more interesting case arises for $g = (-1)^F$, where $F$ denotes spacetime fermion number. Compactifying Type IIA on a circle with $(-1)^F$ holonomy means that fermionic fields are antiperiodic around the circle. We call this an \textit{antiperiodic circle}
\begin{align}
    S^1_A\equiv S^1_{(-1)^F}.
\end{align}
We will refer to the $S^1$ boundary conditions as (P) when fermions are periodic and (A) when antiperiodic.

Under T-duality, this maps to Type IIB orbifolded by $(-1)^F$, namely Type 0B, on a circle with quantum symmetry twist by \eqref{eq:t-duality}. In particular, orbifolding IIB by $(-1)^F$ produces Type 0B.
Furthermore, the quantum symmetry $Q$ associated with the $(-1)^F$ orbifold acts as the left-moving worldsheet GSO parity
\begin{align}
Q = (-1)^{G_L},\end{align}
since it distinguishes the two RR sectors arising from the reversed GSO projection. Consequently, one obtains
\begin{align}
\text{IIA on } S^1_{A}\Big|_{R}
=\text{0B on } S^1_{(-1)^{G_L}}\Big|_{\frac{2}{R}}.
\end{align}
Thus Type IIA compactified with $(-1)^F$ holonomy is T-dual to Type 0B compactified with $(-1)^{G_L}$ holonomy. Taking the zero radius limit $R\to 0$, the 0B side decompactifies to 10d
\begin{align}
    \text{IIA on } S^1_{A}\Big|_{R\to 0}
= \mathrm{0B}.
\end{align}
Another way of getting to 0B in 10d is taking the $R\to 0$ limit of \eqref{eq:0a-0b}
\begin{align}
    \text{0A on } S^1\Big|_{R\to 0}
= \mathrm{0B}.
\end{align}

\section{Difficulties with realizing 0A in M-theory}\label{sec:0A-arguments}
Given that the strong coupling limit of Type IIA is best described by M-theory on a circle,
it is natural to ask what the strong coupling limit of Type 0A theory is and how it may be realized in M-theory.

As we have reviewed in \autoref{sec:T-duality}, Type IIA on an antiperiodic circle $S^1_A$ leads to Type 0B in 10 dimensions in the zero radius limit. 
It is also known that Type 0A on a periodic circle $S^1$ leads to 0B theory in 10 dimensions in the limit of zero radius.  The first path can be viewed as M-theory compactified on $T^2$ with (PA) boundary conditions, where the first entry refers to the 11th direction and the second to the 10th,
\begin{align}
\begin{split}
\text{0B} &=\text{IIA on }S^1_A\Big|_{R\to 0}=\text{M-theory on }T^2_{PA}\Big|_{A\to 0}\\
&= \text{0A on } S^1\Big|_{R\to 0},
\end{split}
\end{align}
where $A$ is the area. If we assume that 0B has an $SL(2,\bZ)$ symmetry and exchange the two circles of M-theory to get (AP) boundary conditions, then identifying 0A with M-theory on the antiperiodic circle $S^1_A$ would be compatible with the T-duality between 0A on a periodic circle $S^1$ and 0B.

This was part of the motivation of \cite{Bergman:1999km} to propose that M-theory on an antiperiodic circle $S^1_A$ leads to 0A.  We now argue this cannot be the case.  We argue against this proposal by three different arguments.  We also argue, as in \cite{Bergman:1999km}, that we cannot expect that 0A theory comes from an orbifold of M-theory by $(-1)^F$ in 11 dimensions compactified on a circle.  

The absence of these two paths from M-theory to 0A forces us to look for unconventional compactifications of M-theory, which we will consider in the next section.

\subsection{M-theory on $S^1_A$ and 0A?}
Let us first argue why M-theory on an antiperiodic circle $S^1_A$ cannot lead to 0A theory in three different ways.

{\bf Argument 1}

The first argument is to note that there are {\it two} distinct orbifolds of IIA involving $(-1)^F$:  
\begin{align}
    g&=(-1)^F\\
    g'&=(-1)^F (-1)^{D_0}
\end{align}
The second action $g'$ is different from the first action $g$ as we are modding out also by the parity of D0-brane charge, which is equivalent in M-theory language to a shift by half the perimeter around the M-theory circle. In other words, orbifolding IIA by $g'$ is equivalent to considering M-theory on an antiperiodic circle
\begin{align}
    \frac{\mathrm{IIA}}{g'} = \text{M-theory on }S^1_A.
\end{align}
If 0A is equivalent to this second orbifold by $g'$, then we learn that $g=g'$.  This is naively impossible, but one could imagine that what we mean by $(-1)^F$ orbifold in IIA is somehow accompanied secretly by an extra $(-1)^{D0}$ which is invisible to IIA perturbation theory, as all IIA states have zero $D0$ charge.  However, that $g$ is distinct from $g'$ is already established in cases where instead of $(-1)^F$ one has a supersymmetric $\bZ_2$ orbifold action like $T^4/\bZ_2$, as studied in \cite{Schwarz:1995bj}.  
The extra $(-1)^{D0}$ action in this case gets rid of all the massless states of the twisted sector.  

However, one could imagine that perhaps the non-supersymmetric version is always accompanied by this additional shift $(-1)^{D0}$ when the spacetime action is $(-1)^F$.  We now argue that this is not possible by considering a non-abelian orbifold which cannot have such an M-theory lift.

Consider an orbifold group $G$. For an orbifold to have such a $\bZ_2$ shift, we have to be able to find a non-trivial one-dimensional representation of the group $G$ 
\begin{align}
    \delta:G\to U(1)
\end{align}
corresponding to the shift in the 11th direction. We now show that there is a non-abelian $G$ which breaks supersymmetry but does not admit a non-trivial one-dimensional representation given by phases, showing that this cannot be realized at least in this example.

Consider $G$ to be the order $60$ icosahedral group, and consider the IIA orbifold $\bR^3/G$. 
Note that since this includes rotations by $\pi$, the square of such elements will have to be doubled to $\hat G\subset SU(2)$ to act properly on fermions as it includes $(-1)^F$ sectors.  On the other hand, $\hat G$ (corresponding to $E_8$ in the McKay correspondence) has no non-trivial one-dimensional representation, which means this $(-1)^F$ sector cannot be lifted to M-theory by including an additional shift of the M-theory circle.  Thus $g$ and $g'$ cannot be equivalent.

{\bf Argument 2}

Type IIA on an antiperiodic circle of radius $R_A$ as $R_A\rightarrow 0$ leads to 0B theory in 10d.
On the other hand, as discussed in \autoref{sec:T-duality}, to get back to IIA from 0B we have to put it on a circle of radius $R_B=2/R_A$ and turn on a $\bZ_2$ quantum symmetry holonomy $Q$ as we go around the circle.
From the M-theory perspective this corresponds to a (PA) spin structure for a torus of radii $(R_1,R_2)$ and their ratio maps to the 0B coupling $\lambda_B=R_1/R_2$.  From the 0B perspective, if we fix $R_B$ to be finite, we need to turn on the quantum symmetry $Q$ around the $R_B$ circle for it to correspond to M-theory on (PA) torus.
Consider taking the coupling to be large $\lambda_B \gg 1$.  This is equivalent to M-theory on $T^2$ with (AP) boundary conditions, where the first entry corresponds to the smaller circle.  If M-theory on (A) is 0A, then this is just 0B on the circle {\it without} holonomy. But we already know that 0B has a $\bZ_2$ quantum symmetry holonomy turned on! 

This would have implied that at strong coupling turning on this holonomy or not does not matter, i.e., that $Q$ is not acting on any perturbative states.  In other words, all the $\bZ_2$ odd $Q$ states of 0B should have become massive by some mechanism, and replaced by new massless ones for which $Q$ acts trivially.  That sounds rather strange to expect.

{\bf Argument 3}

One of the motivations for identifying M-theory on an antiperiodic circle with 0A was based on the expectation that M-theory on $S^1_A$ should lead to a string theory and there was no other candidate dual besides 0A.
Here we explain why we cannot necessarily expect M-theory on an antiperiodic circle to lead to a fundamental string theory in the limit of shrinking circle. 

Consider an M2-brane wrapped around the antiperiodic circle of radius $R$.  This leads to a string which has classical tension $T\sim R M_p^3$.  However, there will also be a Casimir energy contribution to the string tension. Because the $(-1)^F$ twist on the M2-brane worldvolume removes fermion zero modes, we expect more bosonic states, leading to a Casimir contribution to the tension proportional to $\sim -1/R^2$. Putting these together gives
\begin{align}
    T\sim M_p^3R-1/R^2.
\end{align}
We thus learn that if $R\lesssim 1/M_p$, the resulting string is a tachyonic string rather than having a positive tension. 

For odd winding number $N$, this leads to tension $M_p^3 (NR)-1/(NR)^2$, leading to new instabilities at each $(M_pR)\sim 1/N$.  Thus as $R\rightarrow 0$ we get infinitely many instability modes. Note that this is similar to how tachyonic particles arise in Scherk-Schwarz compactification of string theory:  consider wrapping a string $n$ odd times around an antiperiodic circle.  The resulting mass for the particle would be expected to be $nRM_s^2-1/nR$ leading to an instability at $(nR)^2=1/M_s^2$. This is consistent with expectations from perturbative strings which lead to a new tachyon for each odd $n$.

Thus as the radius of M-theory on antiperiodic circle shrinks $R\rightarrow 0$, we cannot expect to get a small positive tension string theory as would be the case if the wrapped M2-brane on the M-theory circle led to the Type 0A string. 

\subsection{M-theory with $(-1)^F$ orbifold and 0A?}
It is natural to ask whether the strong coupling limit of 0A theory is the 11-dimensional M-theory orbifolded by $(-1)^F$.
One argument that this is unlikely was already given in \cite{Bergman:1999km}, as no 11d multiplet on a circle would lead to the one expected in 0A.  One could have imagined that interpolating to zero radius may change the story by Higgsing some fields or getting additional light fields as we shrink the circle. This would be a strange possibility, and we now give another argument why this cannot be the case.

Consider IIA strings on an orbifold $\mathbb{R}^{10}/G$. The partition function of this theory is given by a sum of IIA worldsheet over the torus with $G$-twisted boundary conditions around the worldsheet circles with commuting pairs
    \begin{align}
\sum_{g,h}\farsquare{g}{h}\,.
\end{align}
From the M-theory perspective this would look like an M2-brane with $T^3$ worldvolume where one of the circles of $T^3$ is wrapped once around the 11th circle.  We would also need to include sectors corresponding to the product of the shift along the 11th direction and the $G$ action, as we do in usual orbifolds, otherwise this would not be consistent with the diffeomorphism invariance of M-theory since those twists are allowed in the other directions.  So we should also have sectors (with commuting pairs)\footnote{It is enough to see that $(g,h,1)$ appears in the sum by string perturbation theory, therefore $(1,g,h)$ would appear too, so there would be sectors with the 11th direction twisted.}
\begin{align}
   \sum_{g,h,k}\cubeBC{g}{h}{k}\,.
\end{align}
But we do not have any extra sector contributions in the perturbative string theory!  So for this to be consistent it must be that the M-theory circle twisted by $k\not= 1$ would not contribute perturbatively. Indeed if $k\in G$ acts non-trivially on spacetime, twisting the small circle of M-theory would lead to a large action, and thus a suppression of the form 
\begin{align}
    \exp(-R\cdot (1/M_pR^2))\sim \exp(-1/M_pR)\ll 1
\end{align}
and this can be ignored in the perturbative limit.  This would explain why every known Type IIA orbifold that lifts to M-theory has all orbifold elements acting non-trivially on spacetime.\footnote{Though this may not be a sufficient condition for being liftable.} However, if the orbifold group has any elements that do not act on spacetime, like $(-1)^F$, there is no reason to expect it to be liftable to M-theory.  A simple example is modding out IIA by $(-1)^{F_L}$ which gives IIB theory. Indeed we do not have an 11-dimensional lift of IIB theory, consistent with this no-go expectation. Similarly, we cannot expect the strong coupling limit of 0A to be 11d M-theory with a $(-1)^F$ orbifold.  
This of course does not imply by itself that M-theory does not admit a $(-1)^F$ orbifold, only that if it existed, its compactification on a circle has no reason to be related to 0A.

\section{M-theory origin of Type 0A strings}\label{sec:S1vS1}
In this section we propose a geometric origin of 0A at weak coupling in terms of 11-dimensional M-theory.

\subsection{The proposal}

Having seen that the most obvious connections between 0A and M-theory do not work, it is natural to ask if there are any other, perhaps less conventional ways to connect them.  This is the main aim of this paper and in this section we make such a proposal.

As we saw in the previous section, 0A has two $U(1)$ gauge fields, which we can denote by $A^{\pm}$.  In Type IIA, the existence of a single $U(1)$ gauge field $A$ signals the existence of a circle in M-theory.
The $A$-field of type IIA, which is related to the existence of the $S^1$ in M-theory compactification can be identified with the diagonal combination (coming from the untwisted sector of 0A)
\begin{align}
    A= A^++A^-.
\end{align}
Therefore it is natural to look for the two $U(1)$ gauge fields of 0A to come from two $S^1$'s in the compactifications of M-theory.  But naively this is impossible, because M-theory is an 11-dimensional theory, and there is no room for having the product $S^1\times S^1$ down to 10 dimensions.  One possibility is simply to give up on an M-theory realization.  
However, if we wish to have two $S^1$'s in 11d there is another way:  $S^1\cup S^1$.
But this is a disconnected space, so it cannot lead to a connected quantum theory of gravity.  It turns out there is a resolution of this puzzle if we broaden M-theory compactifications to allow a new exotic class. Consider instead the wedge of two circles: $S^1\vee S^1$, which is two circles connected at a point $p$ like the figure `8' or `$\infty$'.   Of course this is a singular space, as the tangent space at the point the two circles touch is not well-defined. 
One issue to overcome is that $U(1)\times U(1)$ moves the joining point, and so to have a field space invariant under that we should allow the joining point to be allowed to `move' along the two circles. This may make sense, at least when the circles are sub-Planckian where quantum effects may lead to novel extensions of what geometries are allowed.
In addition to defining the theory, we need to decide whether we require fields on the two circles to agree on the joining point, and if so what derivatives mean at this point.

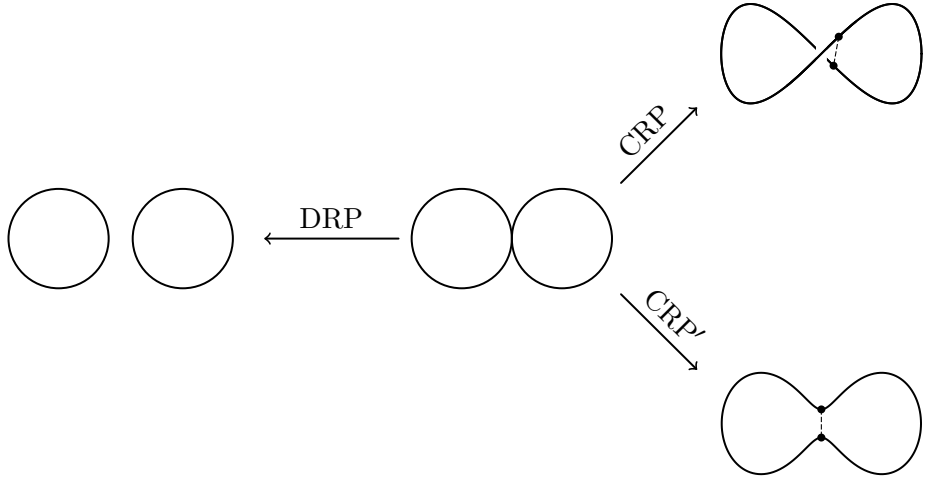
\begin{figure}
\begin{center}
\tikzset{every picture/.style={line width=0.75pt}}

\begin{tikzpicture}[x=0.75pt,y=0.75pt,yscale=1,xscale=1]

\tikzset{
  mydraw/.style={draw,line width=0.75pt,line cap=round,line join=round},
  arr/.style={->,line width=0.75pt,line cap=round}
}

\draw[mydraw] (310.5,0) circle (25);
\draw[mydraw] (360.5,0) circle (25);

\draw[mydraw] (109.25,0) circle (25);
\draw[mydraw] (171.25,0) circle (25);

\pgfdeclarelayer{bg}
\pgfdeclarelayer{fg}
\pgfsetlayers{bg,main,fg}

\def\a{50}   
\def\b{50}   
\def\cx{490}
\def\cy{93}

\begin{pgfonlayer}{bg}
\draw[mydraw]
  plot[domain=0:360,samples=300]
    ({\cx + \a*cos(\x)},
     {\cy + \b*sin(\x)*cos(\x)});
\end{pgfonlayer}

\draw[white,line width=4pt,line cap=round]
  (\cx,\cy-10) -- (\cx,\cy+10);

\begin{pgfonlayer}{fg}
\draw[mydraw]
  plot[domain=70:110,samples=120]
    ({\cx + \a*cos(\x)},
     {\cy + \b*sin(\x)*cos(\x)});
\end{pgfonlayer}

\begin{pgfonlayer}{bg}
\draw[mydraw]
  plot[domain=0:360,samples=300]
    ({\cx + \a*cos(\x)},
     {\cy + \b*sin(\x)*cos(\x)});
\end{pgfonlayer}

\draw[white,line width=4pt,line cap=round]
  (\cx,\cy-10) -- (\cx,\cy+10);

\begin{pgfonlayer}{fg}
\draw[mydraw]
  plot[domain=70:110,samples=120]
    ({\cx + \a*cos(\x)},
     {\cy + \b*sin(\x)*cos(\x)});
\end{pgfonlayer}

\draw[mydraw]
  plot[domain=0:360,samples=300]
    ({\cx + 35 * cos(\x)*sqrt(1^2*cos(2*\x)+sqrt(1.01^4-1^4 *sin(2*\x)^2))},
     {-\cy + \a * sin(\x)*sqrt(1^2*cos(2*\x)+sqrt(1.01^4-1^4 *sin(2*\x)^2))});
\def\t{80}

\coordinate (idUpA) at
({\cx + \a*cos(\t)},
 {\cy + \b*sin(\t)*cos(\t)});

\coordinate (idUpB) at
({\cx + \a*cos(\t+3)},
 {\cy - \b*sin(\t+3)*cos(\t+3)});

\draw[dash pattern=on 2pt off 0.8pt] (idUpA) -- (idUpB);
\fill (idUpA) circle (1.5pt);
\fill (idUpB) circle (1.5pt);
\def\t{90}

\coordinate (idA) at
({\cx + 35*cos(\t)*sqrt(cos(2*\t)+sqrt(1.01^4-sin(2*\t)^2))},
 {-\cy + \a*sin(\t)*sqrt(cos(2*\t)+sqrt(1.01^4-sin(2*\t)^2))});

\coordinate (idB) at
({\cx + 35*cos(\t)*sqrt(cos(2*\t)+sqrt(1.01^4-sin(2*\t)^2))},
 {-\cy - \a*sin(\t)*sqrt(cos(2*\t)+sqrt(1.01^4-sin(2*\t)^2))});

\draw[dash pattern=on 2pt off 0.8pt] (idA) -- (idB);
\fill (idA) circle (1.5pt);
\fill (idB) circle (1.5pt);


\draw[arr] (278.96,0) -- node[midway, above] {DRP} (211.85,0);
\draw[arr] (390, 28) -- node[midway, sloped, above] {CRP} (428,66);
\draw[arr] (390, -28) -- node[midway, sloped, above] {CRP$'$} (428,-66);

\end{tikzpicture}
\end{center}
\caption{Disconnected resolution property (DRP) corresponds to resolving the singularity by disconnecting the two circles. Connected resolution property (CRP and CRP$'$) corresponds to resolving to the double cover. For CRP and CRP$'$ we further impose that the values of the fields on some pair of points corresponding to the resolved point agree. Fields that are smooth under all DRP and CRP and CRP$'$ will be said to have strong smoothness property SSP.}
\label{fig:CRP}
\end{figure}

One way to define this is to consider a resolution of $S^1\vee S^1$.  There are two distinct resolutions and we can consider fields being well-defined on the resolution:
\begin{align}
    S^1\vee S^1&\rightarrow S^1\cup S^1,\\
    S^1\vee S^1 &\rightarrow S^1.
\end{align}
In other words the fields can satisfy the first one and have a \textit{disconnected resolution property} (DRP) and the second the \textit{connected resolution property} (CRP) (see \autoref{fig:CRP}). For CRP, we require the fields to agree at the contact point, whereas for DRP we do not have any contact points and thus no condition. Note that in addition for CRP we can have two distinct resolutions depending on the relative orientation of the two joining circles. Whenever we say that a field satisfies CRP, we mean that there is a sector where the field satisfies CRP and another sector (possibly with flipped chirality) where it satisfies the orientation reversed version CRP$'$, and both are allowed in the spectrum.
To guarantee that we have $U(1)\times U(1)$ symmetry, one would naively think that the only possibility is to require disconnected resolution property (DRP), because the connected resolution property will have a contact point which break the independent $U(1)$ rotations. But if all fields satisfy DRP we would get two copies of the theory that do not interact with one another.  We need some fields to have CRP.  For fields satisfying CRP we restore this symmetry by allowing the joining point to be {\it any} pair of points on the two circles.
 In other words the function space should identify
\begin{align}
    [f(\theta^+),g(\theta^-)]\sim [f(\theta^++\alpha), g(\theta^-\pm\beta)],
\end{align}
where $\alpha, \beta$ are $U(1)\times U(1)$ gauge parameters and $\pm$ above represents the two distinct CRP orientations.  Moreover for the $U(1)$ actions to be compatible with an integral spectrum we need to require that the smoothness of CRP fields is with respect to angular variables $\theta^\pm$ on each circle, and not the length variable.  In other words CRP field space will not depend on the relative length of the two circles.  Fields satisfying both CRP and DRP will be said to satisfy the \textit{strong smoothness property} (SSP).
In this work we will impose fields to satisfy either SSP or DRP. Similarly to CRP, whenever a field satisfies SSP we also allow its orientation reversed version SSP$'$ (possibly with flipped chirality) in the spectrum as well. The requirement that SSP fields admit smoothness under both orientations reflects the fact 0A can be obtained in two ways, from IIA or IIA$'$. In particular, the orientation flip on one of the circles of $S^1\vee S^1$ can be identified with the action of $\Omega$, as we will show, and so both orientations should be allowed in the resolution.

Note that for fields satisfying SSP for equal radii $R$, the first excited KK states will have mass $2/R$ represented by pairs of modes $(\cos \theta^+,\cos \theta^-)$ for the contact point $\theta^+=\theta^-=0$ (and all other combinations represented by independent shifts of $\theta^+\mapsto \theta^++\alpha$ and $\theta^-\mapsto \theta^-\pm \beta$) leading to KK momenta 
\begin{align}
    n[(+1,+1),(+1,-1),(-1,+1),(-1,-1)].
\end{align}
In other words this would allow equal or opposite sign KK momenta on the two circles depending on which CRP resolution they are compatible with. 
On the other hand for DRP fields the KK excitations start with mass $1/R$. Therefore for DRP fields we have KK modes 
\begin{align}
    n[(+1,0), (0,+1), (-1,0), (0,-1)].
\end{align}
See \autoref{fig:wedgeKK} and \autoref{fig:intervalKK} for illustrations of KK modes satisfying DRP and CRP. We will see evidence of this structure when we revisit the $D0$-branes of 0A.

Let us see how we can obtain the full content of Type 0A from M-theory on $S^1\vee S^1$ starting with the metric.  Consider the metric of the 11-dimensional space  $M\times (S^1\vee S^1)$. Let $x$ denote the coordinates of $M$ and $\theta^+,\theta^-$ the coordinates of the two circles.  For the functions which correspond to the connected resolution we get only one circle, whose argument we denote by $\theta^{\pm}$. Since we want a well-defined unique metric on $M$ we must have $g_{\mu\nu}$ in the 10-dimensional directions to be a unique function.
Let us first assume the metric does not depend on the circle coordinates $\theta^\pm$. Then the metric would be parameterized by
\begin{align}
    \begin{split}
        ds^2&=g_{\mu \nu}(x) dx^\mu dx^{\nu}+A_{\mu +}(x)\ dx^\mu d\theta^++A_{\mu -}(x)\ dx^\mu d\theta^-\\
&\quad+R_{+}^2(x)d\theta^{+2}+R_-^2(x) d\theta^{-2}.
    \end{split}
\end{align}
We already note that the symmetric rank-0, 1, 2 tensors (scalars, vectors and metric) of 0A theory are encoded in the metric and equivalent to defining this metric!  Namely
$g_{\mu\nu}$ is identified with the 0A metric, and $A_{\mu\pm}$ with the two gauge fields, and there are two scalars.

\begin{figure}[t]
\centering
\begin{tikzpicture}[scale=1.2]

\def\R{1.6}
\def\Rp{1}
\def\A{0.18}

\draw[thick] (-\R,0) circle (\R);
\draw[thick] ( \Rp,0) circle (\Rp);

\fill (0,0) circle (2pt);
\node[above] at (-0.2,-0.2) {$p$};

\node at (-\R,2.1) {$\theta^+$};
\node at (\Rp,1.4) {$\theta^-$};

\draw[blue,thick,domain=0:360,samples=200]
plot ({-\R + (\R+\A*sin(7*\x))*cos(\x)},
      {      (\R+\A*sin(7*\x))*sin(\x)});

\node[blue] at (-3.0,1.4) {$(n,0)$};

\draw[red,thick,domain=0:360,samples=200]
plot ({ \Rp + (\Rp+\A*sin(7*\x))*cos(\x)},
      {      (\Rp+\A*sin(7*\x))*sin(\x)});

\node[red] at (1.8,1.3) {$(0,n)$};

\end{tikzpicture}

\caption{KK wave modes on the compactification space $S^1\vee S^1$. 
Modes localized on a single circle correspond to DRP fields with KK momenta $(n,0)$ or $(0,n)$, while correlated modes on both circles correspond to SSP fields with momenta $(n,n)$ or $(n,-n)$.}
\label{fig:wedgeKK}
\end{figure}
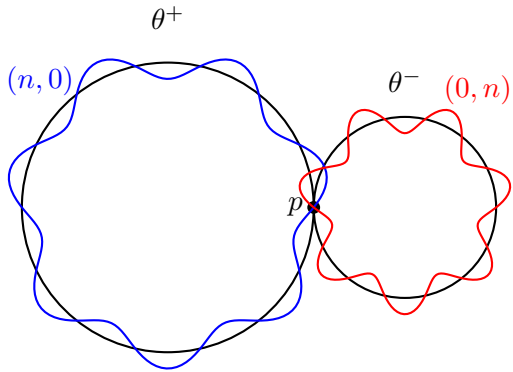
\begin{figure}[t]
\centering
\begin{tikzpicture}[scale=1.2]

\def\L{3}
\def\A{0.35}

\draw[thick] (-\L,2) -- (\L,2);
\draw[dashed] (0,1.6) -- (0,2.4);

\node at (-1.5,2.6) {$\theta^+$};
\node at (1.5,2.6) {$\theta^-$};

\draw[blue,thick,domain=-\L:0,samples=200]
plot (\x,{2 + \A*sin(3*360*(\x+\L)/\L)});

\node[blue] at (-2.6,3.0) {DRP $(n,0)$};

\draw[thick] (-\L,0) -- (\L,0);
\draw[dashed] (0,-0.4) -- (0,0.4);

\node at (-1.5,0.6) {$\theta^+$};
\node at (1.5,0.6) {$\theta^-$};

\draw[red,thick,domain=-\L:\L,samples=200]
plot (\x,{0 + \A*sin(3*360*(\x+\L)/\L)});

\node[red] at (2.5,0.9) {SSP $(n,n)$};

\draw[thick] (-\L,-2) -- (\L,-2);
\draw[dashed] (0,-2.4) -- (0,-1.6);

\node at (-1.5,-1.4) {$\theta^+$};
\node at (1.5,-1.4) {$\theta^-$};

\draw[green!60!black,thick,domain=-\L:0,samples=200]
plot (\x,{-2 + \A*sin(3*360*(\x+\L)/\L)});

\draw[green!60!black,thick,domain=0:\L,samples=200]
plot (\x,{-2 - \A*sin(3*360*(\x+\L)/\L)});

\node[green!60!black] at (2.7,-1.1) {SSP$'$ $(n,-n)$};

\end{tikzpicture}

\caption{KK wave modes on an interval split into regions $\theta^+$ and $\theta^-$. 
DRP modes are localized on one region, while SSP and SSP$'$ correspond to correlated modes across the junction with charges $(n,n)$ and $(n,-n)$. Note that SSP and ${\rm SSP}'$ modes are continuous with respect to DRP as well, i.e. periodic in $\theta^+$ and $\theta^-$ both.}
\label{fig:intervalKK}
\end{figure}
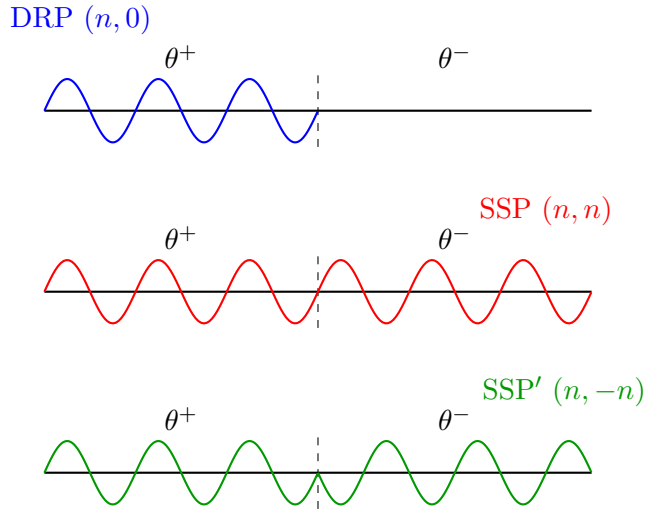

For identifying the scalars with the 0A fields, note that if the circles have equal radii, the geometry has a $\bZ_2$ symmetry exchanging the two circles $S^1\leftrightarrow S^1$. We can consistently identify this with the quantum symmetry action of 0A
\begin{align}
    Q: S^1\leftrightarrow S^1 \qquad A_+ \leftrightarrow A_-.
\end{align}
Using this symmetry, $R^+,R^-$ can be identified with some combinations of dilaton $\phi$ and the tachyon $T$. Indeed taking the evenness of $\phi$ under the $\bZ_2$ symmetry $Q$ leads us to relate it to a symmetric combination of $R$'s such as \begin{align}
    \phi \sim f(R_++R_-),
\end{align} 
leading to a relation expected in IIA context. 
Similarly, $T$ is odd under $Q$ so it can be identified with an antisymmetric combination such as 
\begin{align}
    T\sim ({R_+-R_-})g(R_+,R_-),
\end{align}
with $g$ being a symmetric function of $R_+,R_-$.
Later we make a more precise proposal for both the dilaton and tachyon's relation to the radii.

To have ended up with a single $g_{\mu\nu}$ along $M$ we need to assume the zero mode of $g_{\mu\nu}$ which does not depend on $S^1\vee S^1$ satisfies SSP whereas when $g$ has any component along the circles $A_\pm=g_{\pm *}$ needs to have DRP,
and $R_\pm^2 =g_{\pm\pm}$ being independent should also be DRP fields.
 We can also extend the metric to depend on the circle coordinates in general:
\begin{align}
\begin{split}
    ds^2&=g_{\mu \nu}(x,\theta^\pm)\  dx^\mu dx^{\nu}+A_{\mu +}(x,\theta^+)\ dx^\mu d\theta^++A_{\mu -}(x,\theta^-)\ dx^\mu d\theta^-\\
    &\quad +R_{+}^2(x,\theta^+)d\theta^{+2}+R_-^2(x,\theta^-) d\theta^{-2}.
\end{split}    
\end{align}

To complete the bosonic field content of 0A from this compactification of M-theory we have to make an assignment for the resolution property of various components of the 3-form $C$ field.  We will make the assumption that these have the opposite resolution property to the metric, namely when any component of $C$ is along the circle direction we have SSP and otherwise DRP.  In this way we end up with three fields. A single field when one component is along $S^1\vee S^1$: $C_{\pm \mu \nu} (x,\theta^\pm)$, which will be identified with $B_{\mu \nu}$, and two fields when neither component is: $C_{\mu \nu \rho}^\pm(x,\theta^\pm)$ identified with the two RR 3-form fields.

Under the parity $\theta^\pm\to\pm\theta^\pm$, the DRP fields transform as $C_3^\pm\to\pm C_3^\pm$, while the SSP field $B_{\mu\nu}$ is taken odd under the induced exchange of CRP and CRP$'$. Note that under circle exchange $C^\pm \to C^\mp$, consistent with its identification with quantum symmetry $Q$ earlier.

\begin{table}[t]
\centering
\begin{tabular}{|c|c|c|}
\hline
11d field & 10d field & Resolution property \\
\hline
\hline
$g_{\mu\nu}$ & $g_{\mu\nu}$ & SSP \\
\hline
$g_{\mu\pm}$ & $A_{\mu}^\pm$ & DRP \\
\hline
$g_{\pm\pm}$ & $\phi,\;T$ & DRP \\
\hline
$C_{\pm\mu\nu}$ & $B_{\mu\nu}$ & SSP \\
\hline
$C_{\mu\nu\rho}$ & $C^{\pm}_{\mu\nu\rho}$ & DRP \\
\hline
$\psi_{\mu\alpha}$ & --- & SSP$^*$ \\
\hline
\end{tabular}
\caption{Relation between components of 11d M-theory fields on $S^1\vee S^1$ and the resulting 10d massless fields of Type 0A, together with their assigned resolution properties. The resolution property of the gravitino SSP$^*$ corresponds to SSP with an odd identification $\psi\to -\psi$ at the contact point, removing the massless fermionic zero mode.}
\label{fig:S1vS1-spectrum}
\end{table}

The SSP of $C_{\pm\mu\nu}$ also implies that the M2-brane
along the internal direction should always wrap both circles an equal number of times.  In particular, we obtain only one 10d string by wrapping the M2-brane on the circles and not two.  This was partly our motivation for the SSP when any components of $C$ are along the circles. This implies that if only one of the two circles shrinks we do not get a tensionless string.  We thus expect that the shrinking of only one of these circles is at finite distance in field space, consistent with the distance conjecture \cite{Ooguri:2006in}. This will become relevant when we discuss the tachyon condensation of 0A. 

We have identified the quantum symmetry $Q$ as the circle exchange in M-theory, now we identify geometrically how worldsheet parity symmetry $\Omega$ is realized. As we discussed in \autoref{sec:spectrum} there are two ways to get to 0A theory from orbifolding the Type II theory with $(-1)^F$:  using either IIA or ${\rm IIA}'$. Similarly, the two resolutions CRP and CRP$'$ differ by flipping the orientation of the second circle $\theta^- \to -\theta^-$.  Indeed parity reflection on the second circle acts on the gauge fields by 
 \begin{align}
     \begin{split}
         A^\pm&\rightarrow \pm A^\pm,\\
         C^\pm& \rightarrow \pm C^\pm,
     \end{split}
 \end{align}
 which is exactly the symmetry $\Omega$ in 10d that maps the IIA realization of 0A to that of ${\rm IIA}'$. This completes a full accounting of the symmetries and the bosonic content of 0A from M-theory.

To complete the field content we need to specify how the M-theory gravitino resolves.
We will assume it has SSP, with an `odd structure'.  What we mean by this is that we consider the connected double cover of the circle and when we identify two points on it to get $S^1\vee S^1$, we also identify $\psi_{\mu\alpha}\rightarrow -\psi_{\mu\alpha}$.  In particular this does not lead to any massless gravitino fields as the constant mode is not allowed on $S^1\vee S^1$, consistent with the absence of massless fermions in 0A.  We assume the two distinct gravitino fields under $\Gamma^{11}$ chirality will couple through distinct CRP resolutions.
In particular we assume that the orientation of the two circles agree for the CRP resolution for $\Gamma^{11}=+1$ and opposite orientation CRP$'$ for $\Gamma^{11}=-1$.   For example if $(\sin\theta^+,\sin \theta^-)$ is a $\Gamma^{11}=+1$ gravitino configuration where $\theta^\pm=0$ is the contact point we identify it with shifts $[\sin(\theta^++\alpha),\sin (\theta^-+\beta)]$  made of KK momenta $(+1,+1)\oplus (-1,-1)$ combinations $\frac{1}{2i}[(e^{i\theta^+},e^{i\theta^-})-(e^{-i\theta^+},e^{-i\theta^-})]$, then the $\Gamma^{11}=-1$ configuration could be  obtained by a $\theta^-\rightarrow -\theta^-$ leading to $[\sin(\theta^++\alpha),\sin (-\theta^-+\beta)]$ viewed as combinations of KK momenta $(1,-1)\oplus (-1,1)$.  This implies that Majorana-Weyl spinors from the 10d perspective should come from $n(+1,+1)$ KK momenta (with $n=0$ missing due to vanishing condition at the contact point), whereas conjugate Majorana-Weyl spinors come from $n(1,-1)$ KK momenta around the two circles. See \autoref{fig:S1vS1-spectrum} for a summary of the spectrum.

The fact that different fields of M-theory have distinct resolution properties might at first appear inconsistent.  But this implies that this compactification does not have either resolution allowed classically. So the wedge property of the geometry is pinned corresponding to a frozen singularity.  The only way the singularity can be resolved is if one or the other circle shrinks.  We will argue later that this corresponds to tachyon condensation of 0A leading to IIA theory, as anticipated in \cite{Gutperle:2001mb, Russo:2001tf, David:2001vm, Suyama:2001gd}.

\subsection{The Frozen Singularity and discontinuity of the Lagrangian}
The field space we have defined is `inspired' by the singular space $S^1\vee S^1$. However, it is important to note that there is no classical sense to the space we have.  Rather we have defined the space through the field space.  In particular requiring both DRP for some fields and SSP for others, let alone the fact that the contact point of the circles can be arbitrary, means that a more proper way of thinking about what we have done is to define a quantum version of $S^1\vee S^1$.  This is somewhat analogous to noncommutative manifolds which are defined via the algebra of functions defined on them.  As we will discuss later, it is conceivable that this space makes sense only at sub-Planckian perimeters, consistent with quantum mechanical effects being important in making sense of this geometry in the terms we have proposed.

The fact that some of the fields see the space as disconnected, and some see it as a connected space at different pairs of points means of course that the classical notion of topology should not be sensible and quantum effects should be important to make sense of this compactification.  This is similar to what happens for example for Type IIA strings near a flop transition of a Calabi-Yau.  As we approach the flop there is no unambiguous way to say whether we are on one side of the flop or the other (and we can indeed avoid the singular flop point altogether because the K\"ahler moduli is complexified).  So we can continuously go from one classical geometry to a topologically distinct one, without passing through any singularities.  Of course near the singular point it is maximally ambiguous to say on which manifold the string is propagating, i.e. whether we are on one side of the flop or the other. In our case it is as if we are frozen near the singular point and we cannot move away from it.  We are conjecturing that similar quantum effects are responsible for making the inconsistent classical geometry seen by different resolutions of fields to actually end up being consistent in the full quantum theory.\footnote{One might wonder whether the quantum version of $S^1\vee S^1$ would admit a perturbative worldsheet description that would resolve the singularity. However, this may not be possible. Strings wind around $S^1\vee S^1$ according to CRP, but some fields obey DRP. Consequently perturbative string oscillations may not naturally detect the DRP fields. This is not unusual: for example at the conifold point perturbative string theory also becomes inadequate due to additional massless states that are not visible to perturbative strings. Therefore the lack of worldsheet description is not a threat to consistency.}

We thus conjecture that quantum effects will render this frozen singularity a consistent background for M-theory (at least for sub-Planckian circles) with different fields behaving as if they live on topologically distinct spaces.

\subsection{Further evidence}

It is quite satisfying that the full field content of 0A can be read off from the geometric description of M-theory compactified on $S^{1+}\vee S^{1-}$.  Note that in particular each of the doubled fields of 0A corresponds to fields living on one or the other connected smooth space of the form $M\times S^{1\pm}$.  Note also that the $\bZ_2$ quantum symmetry of 0A exchanges $S^{1+}\leftrightarrow S^{1-}$ and the corresponding fields.  

\subsubsection{$D0^{\pm}$-branes and KK modes}
There is further evidence for this picture coming from the expected KK modes.
The M-theory picture on $S^1\vee S^1$ should become better as we go to large coupling of 0A theory as is the case in the context of IIA theory.  However, unlike the IIA theory where supersymmetry guaranteed that the KK modes of supergravity fields form mass degenerated supersymmetric multiplet with mass exactly given by $n/R$, for the 0A theory the multiplets can have different masses.  So there is no clear picture that in the limit of small circles where the supergravity description is inappropriate of what happens.  Nevertheless we can still make some robust predictions.

The fact that there are two circles (related to the two $U(1)$ gauge fields of 0A) implies that there must exist stable KK momenta. We therefore have to have at least one of the supergravity modes to have the lightest charged state as $R\rightarrow 0$ with charges $(1,0)$ and $(0,1)$.  The most natural candidate to expect to be the lightest KK modes would be the scalar fields $R_+\sim \exp(i \theta^+)$ carrying $(1,0)$ charge and $R_-\sim \exp(i\theta^-)$ with charge $(0,1)$.  The other bosonic KK multiplets can in principle be unstable and decay to the massless fields which carry spacetime vectorial quantum numbers and these KK modes. In addition M-theory does have spinors, so the quantum number of spinors should be possible to be observed as part of KK spectrum.
The lightest states carrying spinor charge must come from the gravitino field and given what we discussed the lightest such state must carry $U(1)\times U(1)$ charges $(1,1)$ for 10d MW spinors and $(1,-1)$ for the conjugate ones.

We now provide evidence that these expectations are indeed supported by studying the D0-branes of 0A theory.

As we have reviewed in \autoref{sec:branes}, 0A theory has two 0-branes: $D0^\pm$.  This would naturally correspond to momenta around the corresponding two circles.  
Indeed the worldvolume of each simply gives 9 bosonic fields which correspond to moving the KK momentum mode along the transverse space to the circles.  Moreover as we discussed the masses of these branes vary with the vev of the tachyon and the dilaton to leading order as
\begin{align}
    m_{D0^\pm}= \frac{1}{\sqrt{2}}\frac{M_s}{\lambda_s}\left(1\mp {T\over 4}+\mathcal O(T^2)\right).
\end{align}
This matches the picture we have.  To leading order, the tachyon field is proportional to the difference between the two radii.  Since the masses of KK modes are proportional to inverse radii when the tachyon vev is zero they should be equal, which is the case here.  Moreover as we vary the radii, the KK modes of the two circles will have different masses which to first order satisfy
\begin{align}
    \frac{\delta m}{m}\sim \frac{\delta R}{R},
\end{align}
which matches the expected mass of $D0^\pm$ if we assume that to leading order $T\propto \frac{\delta R}{R}$.
Our proposal predicts that the KK gravitons coming from the individual circles can be scalar fields corresponding to the leading KK excitations of $R_\pm$. This is consistent with the fact that in the worldvolume of $D0^\pm$-branes we do not have massless fermions and it is purely bosonic and has spin 0.  Note that the fact that we have no fermions on $D0^\pm$ to quantize to give the Clifford algebra leading to higher spins is consistent with the lack of expected degeneracy for KK modes.  In particular the KK modes of $A_\mu^{\pm}$ around each circle are presumably heavier and decay to combinations of massless fields and the scalar KK modes just discussed.

As we have noted, we expect there are fermionic states with charges $(1,1)$ or $(1,-1)$ with opposite 10d chirality.  We thus consider 0A theory in the presence of coincident $D0^+-D0^-$ and
$D0^+-\overline{D0}^-$ pairs.
 Indeed as we have discussed in \autoref{sec:branes} this does lead for both pairs of 0-branes to massless fermions on the worldvolume of the open string stretched between the pair---MW spinor or conjugate spinor depending on which pair we consider.  This suggests a bound state of the 0-branes would indeed realize the fermions we have expected from the gravitino discussion above. There could also be other bound states in these combinations for other fields satisfying SSP including those for $g_{\mu \nu}$ and $B_{\mu\nu}$, but this is not a necessary requirement for our picture to hold as they could be more massive.

\subsubsection{IIB from 0A}

Further evidence for our proposal comes from the fact that if we place the 0A theory on a circle with a quantum $\bZ_2$ holonomy around the circle it should lead in the small radius limit to the IIB theory in 10d, using the result of \autoref{app:orbifold}.  

According to our proposal, M-theory on $S^1\vee S^1$ leads to 0A.  If we put this on another circle with the quantum $\bZ_2$ symmetry, and take this to zero radius, it means that we are projecting to the $\bZ_2$ invariant sector in the limit.  Since the $\bZ_2$ symmetry acts by exchanging the two circles of $S^1\vee S^1$, invariant subspace means that the theory is fully described by fields on only one of the $S^1$'s.  Combined with the extra circle, we get that the zero radius limit of 0A with a $\bZ_2$ quantum symmetry holonomy should be the same as M-theory on $T^2$ which is indeed the F-theory picture for Type IIB theory!\footnote{Note that the orbifold of 0A not acting on spacetime have no reason to be liftable to M-theory as we discussed in \autoref{sec:0A-arguments}.  If we orbifold 0A in 10d by quantum symmetry, this leads to IIA.  There is no reason there should be an M-theory lift of the orbifold, but nevertheless it still seems to work as the quantum symmetry is exchanging the two circles leaving us with one circle.}

\subsubsection{Tachyon condensation}

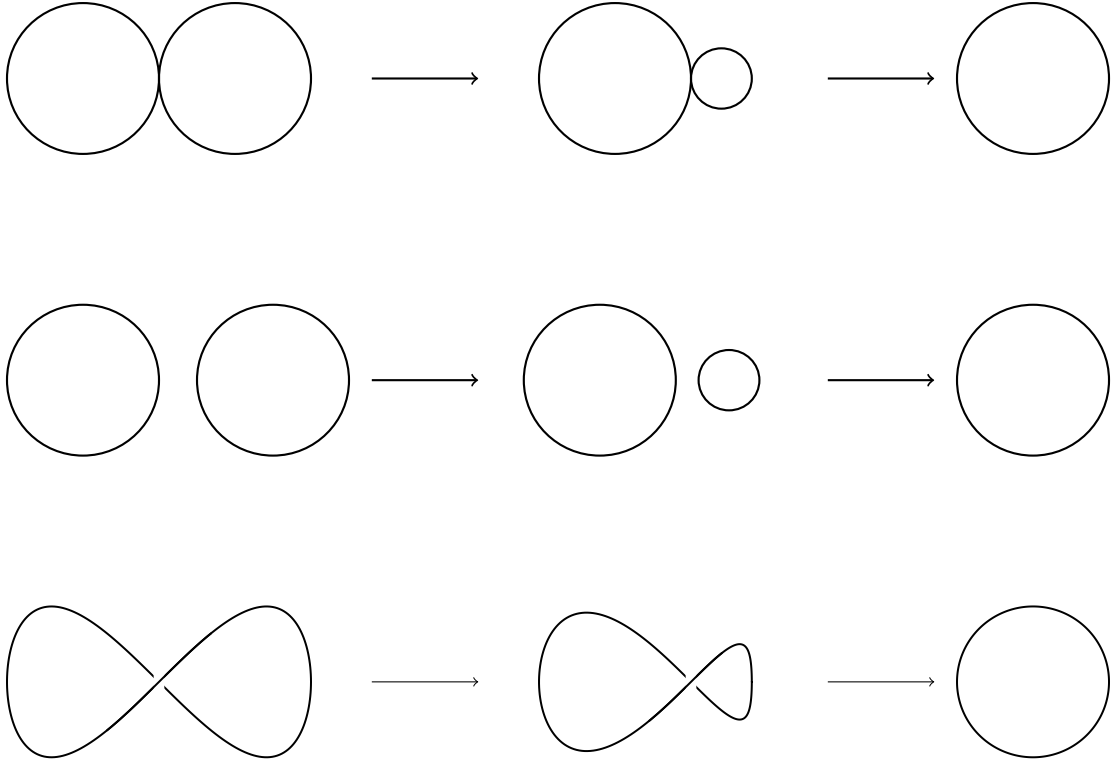
\begin{figure}
    \centering
    \begin{tikzpicture}[thick]
\pgfdeclarelayer{bg}
\pgfdeclarelayer{fg}
\pgfsetlayers{bg,main,fg}

\tikzset{
  mydraw/.style={draw,line width=0.75pt,line cap=round,line join=round},
  arr/.style={->,line width=0.75pt,line cap=round}
}

\def\xA{0}
\def\xB{2}
\def\xArrowOneStart{3.8}
\def\xArrowOneEnd{5.2}
\def\xC{6.5}
\def\xArrowTwoStart{9.8}
\def\xArrowTwoEnd{11.2}
\def\xE{12.5}

\draw (\xA,0) circle (1);
\draw ({\xA+2},0) circle (1);

\draw[->] (\xArrowOneStart,0) -- (\xArrowOneEnd,0);

\draw (\xC+0.5,0) circle (1);
\draw ({\xC+1.9},0) circle (0.4); 

\draw[->] (\xArrowTwoStart,0) -- (\xArrowTwoEnd,0);

\draw (\xE,0) circle (1);

\draw (\xA,-4) circle (1);
\draw ({\xA+2.5},-4) circle (1);

\draw[->] (\xArrowOneStart,-4) -- (\xArrowOneEnd,-4);

\draw (\xC+0.3,-4) circle (1);
\draw ({\xC+2.0},-4) circle (0.4);

\draw[->] (\xArrowTwoStart,-4) -- (\xArrowTwoEnd,-4);

\draw (\xE,-4) circle (1);


\def\cy{-8}
\def\R{1}      
\def\a{2}     
\def\b{2}      
\def\cut{0.15}

\def\cx{\xA+1}
\def\eps{0}

\begin{pgfonlayer}{bg}
\draw[mydraw]
  plot[domain=0:360,samples=260]
    ({\cx + (\a*(1-\eps*(1+cos(\x))/2))*cos(\x)},
     {\cy + (\b*(1-\eps*(1+cos(\x))/2))*sin(\x)*cos(\x)});
\end{pgfonlayer}

\draw[white,line width=4pt,line cap=round]
  (\cx,\cy-\cut) -- (\cx,\cy+\cut);

\begin{pgfonlayer}{fg}
\draw[mydraw]
  plot[domain=70:110,samples=120]
    ({\cx + (\a*(1-\eps*(1+cos(\x))/2))*cos(\x)},
     {\cy + (\b*(1-\eps*(1+cos(\x))/2))*sin(\x)*cos(\x)});
\end{pgfonlayer}

\draw[->] (\xArrowOneStart,\cy) -- (\xArrowOneEnd,\cy);

\def\cx{\xC+1.5}
\def\eps{0.6}

\begin{pgfonlayer}{bg}
\draw[mydraw]
  plot[domain=0:360,samples=260]
    ({\cx + (\a*(1-\eps*(1+cos(\x))/2))*cos(\x)},
     {\cy + (\b*(1-\eps*(1+cos(\x))/2))*sin(\x)*cos(\x)});
\end{pgfonlayer}

\draw[white,line width=4pt,line cap=round]
  (\cx,\cy-\cut) -- (\cx,\cy+\cut);

\begin{pgfonlayer}{fg}
\draw[mydraw]
  plot[domain=70:110,samples=120]
    ({\cx + (\a*(1-\eps*(1+cos(\x))/2))*cos(\x)},
     {\cy + (\b*(1-\eps*(1+cos(\x))/2))*sin(\x)*cos(\x)});
\end{pgfonlayer}

\draw[->] (\xArrowTwoStart,\cy) -- (\xArrowTwoEnd,\cy);

\def\cx{\xE}

\draw[mydraw]
  plot[domain=0:360,samples=260]
    ({\cx + \R*cos(\x)},
     {\cy + \R*sin(\x)});
\end{tikzpicture}          
    \caption{Tachyon condensation of 0A corresponds to shrinking of one of the circles, which leaves only one circle at the end, which is IIA, i.e., M-theory on $S^1$.}
    \label{fig:tachyon}
\end{figure}
From the 0A perspective we know that the tachyon field has an instability which should lead to a more stable vacuum by condensing.  Given our identification of the tachyon as a measure of the difference of the radii of the two circles would geometrically correspond to shrinking one circle relative to the other.  Indeed if we completely shrink one of the circles we literally are back with M-theory on a circle which is IIA theory, see \autoref{fig:tachyon}.  This leads to the prediction that the end result of 0A tachyon condensation is IIA theory.  It is important to note as we have already mentioned that tachyon condensation is at finite distance in field space, because as we shrink one circle we do not end with a tower of light states:  the KK modes become very heavy, and the winding membrane wraps on pairs of circles not individual ones.  Therefore the distance conjecture would be compatible with this happening at finite distance in tachyon field space.  This is in agreement with other arguments expecting that IIA would be the end result of tachyon condensation \cite{Gutperle:2001mb, Russo:2001tf, David:2001vm, Suyama:2001gd}.

If our picture is correct we would be predicting that at some finite vev of the tachyon field the mass of $D0^-$ should go to infinity.  The mass of $D0^{\pm}$ has been computed to second order of tachyon vev \cite{Garousi:1999fu}, with the result
\begin{align}
    m_{D0^\pm}=\frac{M_s}{\sqrt{2}\lambda_s}\left(1\mp \frac{T}{4}+\frac{3T^2}{32}+...\right).
\end{align}
Based on this, Garousi has conjectured in \cite{Garousi:1999fu} that the full expression is given by 
\begin{align}
    m_{D0^\pm}=\frac{M_s}{\sqrt{2}\cdot \lambda_s\sqrt{1\pm\frac{T}{2}}}.
\end{align}
This conjecture fits beautifully with our prediction!  If we set the vev $\langle \frac{T}{2}\rangle =1$ it leads to 
\begin{align}
    m_{D0^-}=\infty,\quad  m_{D0^+}=\frac{M_s}{2 \lambda_s},
\end{align}
and this is consistent with the expectation that as the second circle shrinks the mass of $D0^-$ corresponding to the KK mode should go to infinity, while the mass of $D0^+$ should remain finite, and this is at finite distance in tachyon field space!  In the M-theory frame this would suggest the relation 
\begin{align}
    R_{\pm}^2=\frac{2\lambda_s^2}{M_s^2}\left(1\pm \frac{T}{2}\right).
\end{align}
Also identifying M2-branes wrapped around $S^1\vee S^1$ with 0A strings we learn
\begin{align}
    M_s^2=M_p^3(R_++R_-),
\end{align}
we are thus led to the relations
\begin{align}
    \lambda_s^2&=\frac{M_p^3}{4}(R_+^2+R_-^2)(R_++R_-),\\
    \frac{T}{2}&=\frac{R_+^2-R_-^2}{R_+^2+R_-^2}.
\end{align}
In the limit of full tachyon condensation (where $R_-=0$) we get IIA theory, which is related to the above parameters by
\begin{align}
    R_A=R_+, \quad \lambda_A=2\lambda_s,
\end{align}
and $D0^+$ becoming the $D0$ of IIA.

\begin{figure}
    \centering

\pgfplotsset{compat=1.18}

\begin{tikzpicture}
\begin{axis}[
view={60}{30},
xlabel={$T$},
ylabel={$\phi$},
zlabel={$V$},
domain=-2:2,
y domain=0:4,
samples=45,
samples y=35,
ytick={0},
ztick={0},
enlargelimits=false
]

\addplot3[
surf,
shader=flat,
draw=black,
fill=green!40
]
{exp(-0.6*y)*(x^2/10 - 1/32)*(x^2 - 2)*(x^2-4)^2};

\addplot3[
very thick,
black,
domain=0:4,
samples=100,
samples y=0
]
({0},{x},{exp(-0.6*x)});
\node at (axis cs:0,1.2,{exp(-0.6*1.4)+0.4}) {$\sim e^{-\frac{5}{\sqrt{2}}\phi}$};

\end{axis}
\end{tikzpicture}
    \caption{Potential in the dilaton–tachyon $(T,\phi)$ field space. Along $T=0$, the dilaton exhibits a runaway in Einstein frame proportional to the real part of \eqref{eq:V0A} as $V\sim e^{-\frac 5 {\sqrt{2}}\,\phi}$. Tachyon condensation to $T=\pm2$ corresponds to shrinking $S^{1\mp}$, leading to the Type IIA vacuum with $V=0$. We use the convention that large $\phi$ is weak coupling.}
    \label{fig:0A-V}
\end{figure}
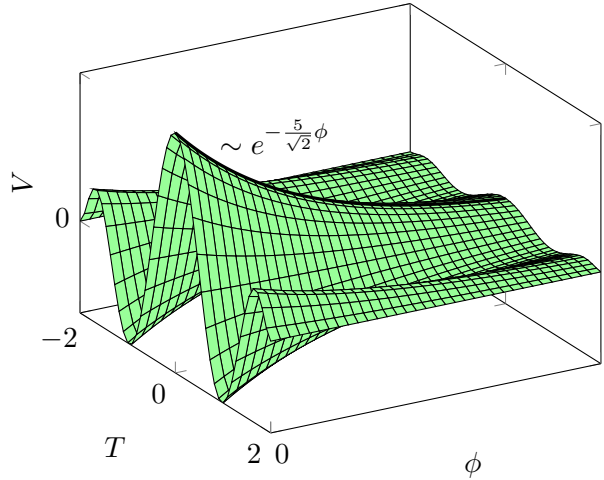

The 0A theory at tree level has vanishing vacuum energy, so rolling on the tachyon which decreases the energy would end up at a negative value, unlike the expected vanishing vacuum energy for IIA theory.  Of course ending up at IIA simply means we would have to come back to $V=0$ as we change the tachyon vev to $T=2$.   One could also ask what the perturbative correction to $V(T=0)$ at 1-loop is. Naively this should be proportional to $V_{\rm tachyon}\propto -m^{10}=-(m^2)^5>0$.  So it could in principle lift the vacuum energy to positive value.  This is indeed the case.  The full one loop vacuum energy for 0A, including the tachyon field contribution as well as all the other fields, was computed by Lorenz Eberhardt upon our request, using the techniques in \cite{Baccianti:2025gll}, with the result in the string frame
\begin{align}\label{eq:V0A}
V_{\rm 0A}=V_{\rm 0B}= - \frac{1}{2}\int_{\mathcal F} \frac{d^2 \tau}{({\rm Im} \tau)^6} \frac{1}{|\eta^{12}|^2}\left(|\vartheta_2^4|^2+|\vartheta_3^4|^2+|\vartheta_4^4|^2\right) \approx 11.36 \pm \frac{4 \pi^6 }{15}i.
\end{align}
The imaginary piece signals the vacuum decay rate per unit volume ($\propto |m_{\rm tachyon}|^{10}$) and the real part signifies the vacuum energy.  It is interesting that this value, $V\sim 11.36\ M_s^{10}$ is positive see \autoref{fig:0A-V}.

It is not clear what will happen when the radii of the circles become large.  If we naively extrapolate the potential, it may suggest when $R_{\pm}\gtrsim  1$, $V\gtrsim 1$ in 11d Planck units which would mean that the theory breaks down and we cannot take $R>1$.  This may be compatible with the fact that our field space assignments which are not fully local would find a natural explanation as quantum effects from the M-theory perspective.

\section{F-theoretic Description of Type 0B and its implications }\label{sec:F-0B}

In this section we use the M-theory realization of 0B strings to develop an F-theoretic description for it.\footnote{For an M and F-theory realization of a non-supersymmetric S self-dual orientifold of Type IIB strings, see \cite{ss4}. }  

\subsection{M-theory and 0B theory}
It is natural to ask what the M-theory lift of 0A tells us about Type 0B.  In the superstring case a useful perspective about symmetries of IIB in 10d comes from the F-theory picture which corresponds to compactifying IIA on a circle and taking the radius to 0 and obtaining IIB in 10d by T-duality.
From the M-theory perspective we have the moduli of Type IIB on a circle related to M-theory on $T^2$, which has two moduli, the complex structure $\tau$ and the area of the torus $A$.  The $\tau$ gets identified with IIB coupling in 10d, and the area gets mapped to $R_B$ by
\begin{align}
    \frac{1}{R_B}=A
\end{align}
where we have taken 11d Planck mass as $M_p=1$. The IIB in 10d corresponds to taking the zero area limit $A\rightarrow 0$ of M-theory with a fixed complex structure $\tau$, leaving us with one complex parameter in 10d which is the axio-dilaton of Type IIB.  The $SL(2,\bZ)$ duality symmetry of IIB follows from geometric duality of M-theory compactification on $T^2$.

We wish to find a similar F-theory picture for 0B.
As we reviewed in \autoref{sec:T-duality}, Type 0B can be obtained by T-duality from Type 0A on a circle and taking the radius to 0. From the M-theory perspective this corresponds to compactification on $(S^{1+}\vee S^{1-})\times S^1$. Let the corresponding radii be
$R_+,R_-,R$. Recall that $\chi_+$ and $\chi_-$ are the axionic fields of 0B from the Ramond sector. Consider tilting the two tori $(R,R_+)$ and $(R,R_-)$ using these shifts:
\begin{align}
    \tau_\pm=i\frac{R}{R_\pm}+\chi_\pm.
\end{align}
Our convention for the choice of $\tau_\pm$ we get from M-theory is that its imaginary component is large as $R\gg R_\pm$.
Under the quantum symmetry $\tau_+\leftrightarrow \tau_-$.  Just as in the 0A case, we identify the 0B axio-dilaton coupling with the symmetric combination and $T$ and $\chi'$ with antisymmetric combinations:
\begin{align}
    {i\over g_s}+\chi &\sim {\tau_+ +\tau_-},\\
    T+i\chi'&\sim \frac{\tau_+-\tau_-}{\tau_++\tau_-}.
\end{align}
Moreover the radius of the circle 0B is compactified upon would be
\begin{align}
    \frac{1}{R_B}=(R_++R_-)R.
\end{align}
Thus the 10d limit is obtained by $(R,R_++R_-)\rightarrow (0,0)$ keeping the ratios of $R/R_{\pm}$ finite.  The F-theory limit of 0B is thus characterized by two complex parameters $\tau_+,\tau_-$.

The natural question is what is the duality symmetry in this case?  Naively one may think there are two $SL(2,\bZ)$ symmetries.  That cannot be the case because switching the 11th and 10th direction acts on both $\tau_{\pm}$ at the same time.  So roughly speaking we have at most one $SL(2,\bZ)$.  We now argue that the symmetry is a level 2 subgroup of $SL(2,\bZ)$. We show this by considering the limit where $\tau_+=\tau_-=\tau$.

It is convenient to consider the $S$-transformed variables $\hat \tau_\pm =-1/ \tau_\pm$.  The DRP fields will see a torus with moduli $\hat \tau=\hat \tau_\pm $ and the SSP fields will see the combined torus moduli $\hat \tau_+ +\hat \tau_-=2\hat \tau$. Since the symmetry acts on the full theory it should be symmetry of both DRP and SSP fields. We thus ask if there is a subgroup of $SL(2,\bZ)$ acting on $\hat \tau$ for which there is a corresponding but possibly distinct $SL(2,\bZ)$ action on the torus where the modulus $2\hat\tau$ also transforms like the modulus of a torus?  We consider
\begin{align}
    \hat \tau&\rightarrow \frac{a\hat \tau +b}{c\hat \tau+d},\\
2\hat\tau&\rightarrow \frac{A(2\hat \tau)+B}{C(2\hat \tau)+D},
\end{align}
and ask how these can be compatible with integral entries when the transformations belong to $SL(2,\bZ)$.
These are compatible if we identify
\begin{align}
    A=a,\quad B=2b,\quad C=\frac{c}{2},\quad D=d,
\end{align}
to get consistent transformations. Note that
\begin{align}
    ad-bc=AD-BC=1.
\end{align}
For this identification to involve integral entries we need $B$ and $c$ to be even.  Going back to the original parameterization $\tau$ we need to conjugate these elements by $S$ transformation which switches the off-diagonal terms.  We thus find that for every element of the level 2 subgroup of $SL(2,\bZ)$ acting on $\tau$ represented by 
\begin{align}
\Gamma^0(2)
&=
\left\{
\begin{pmatrix}
a & c \\
b & d
\end{pmatrix}
\in SL(2,\mathbb{Z})
\;\middle|\;
c \equiv 0 \pmod{2}
\right\},
\end{align} 
we have compatible transformations.  We conclude that the 0B theory on the locus of the $\bZ_2$ quantum symmetric point with $T=\chi'=0$ enjoys $\Gamma^0(2)$ duality symmetry acting on the axio-dilaton $\tau$. We now derive the same result from a totally different perspective using only well established dualities of M-theory.

\subsection{The IIA perspective}
As we discussed in \autoref{sec:T-duality}, another way to end up with 0B in 10d is to consider compactification of the IIA theory on a circle with antiperiodic boundary conditions for fermions and taking its zero radius limit. We can also turn on a $U(1)$ RR holonomy $\theta$ on this circle. In this picture we get a different parameterization of the moduli of the 0B theory.  Namely from the M-theory perspective we are compactifying first on a circle with a periodic boundary condition of radius $R^{11}$ to IIA and then on a circle of radius $R^{10}$ with antiperiodic boundary condition.  This is equivalent to 0B on a circle of radius $R'_B$ with quantum symmetry as the holonomy. The map to the 0B variables is
\begin{align}
    \tau_B&=i\frac{R^{10}}{R^{11}}+\theta,\\
\frac{1}{R'_B}&=R^{11}R^{10},
\end{align}
where we have turned off the tachyon field and the additional RR axion field as $T=\chi'=0$. The tachyon and all the other fields in the twisted sector of 0B with odd momentum along the circle arise in M-theory as odd number of M2-branes wrapped around $T^2$, and so in this sense they are not geometrized, unlike the 0A picture we presented before.

\begin{figure}[t]
\centering
\begin{tikzpicture}[scale=2]

\def\H{2.2}

\begin{scope}
\clip (-1,0) rectangle (1,\H);
\fill[gray!30,even odd rule]
(-1,0) rectangle (1,\H)
(0,0) circle (1);
\end{scope}

\draw[thick] (-1,0) -- (-1,\H);
\draw[thick] (1,0) -- (1,\H);
\draw[thick] (-1,0) arc (180:0:1);

\draw[->] (-1.3,0) -- (1.3,0) node[right] {};
\draw[->] (0,0) -- (0,\H+0.2) node[above] {};

\node at (-1,-0.1) {$-1$};
\node at (1,-0.1) {$1$};

\node[below] at (1+0.25,0+0.5) {$\farsquare{A}{P}$};
\node[above] at (0+0.1,\H+0.05) {$\farsquare{A}{A}$};
\node[above] at (1,\H+0.05) {$\farsquare{P}{A}$};

\fill (0,1) circle (0.03);
\node[left] at (0,1-0.2) {$i$};

\draw[->,thick]
(0.2,1)
arc (0:180:0.18);

\node at (0.1,1.3) {$S$};

\end{tikzpicture}

\caption{
Fundamental domain of $\Gamma_\theta = (TS)\Gamma_0(2)(TS)^{-1}$ for the axio-dilaton $\tau_B' =- \frac{1}{\tau_B}+1 $. 
This parametrization is related to the one used in the IIA perspective by a $TS$ transformation, so the discussion there should be conjugated by $TS$ when interpreting this figure. 
The $\tau_B'$ fundamental domain is particularly simple: the strip $-1\le \mathrm{Re}\,\tau'\le 1$ with the unit disk $|\tau'|<1$ removed. 
The three limits $\tau'_B=1$, $\tau'_B=i\infty$, and $\tau'_B=1+i\infty$ correspond to weak-coupling limits of Type~0B with the three different spin structures, with the latter two belonging to the same weak-coupling regime at different axion values. 
The critical point $\tau_B=(1+i)/2$ corresponds to $\tau_B'=i$, shown in the figure. 
It is fixed by the modular transformation $S$, which acts locally as a $\mathbb{Z}_2$ orbifold. 
}
\label{fig:gamma0}
\end{figure}
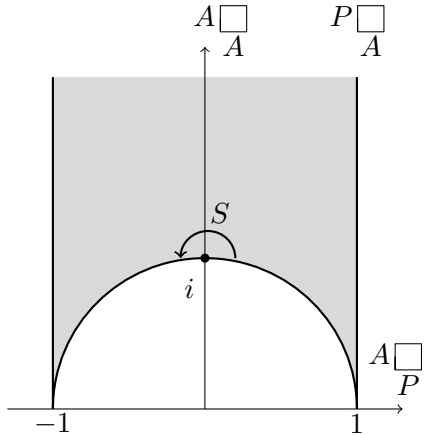

To get the 0B theory in 10d we have to take the zero radius limit of both radii of M-theory with the ratio fixed, as in F-theory, where $\tau$ survives.
We can immediately deduce the duality symmetry acting on $\tau$ from the M-theory perspective of a compactification on torus with a specific spin structure for fermions:  it should be the subgroup of $SL(2,\bZ)$ preserving $T^2_{PA}$ with one circle with periodic boundary condition and the other with antiperiodic boundary condition, and this is precisely $\Gamma_0(2)$! See \autoref{fig:gamma0} for the corresponding fundamental domain.

Note that the IIA compactification on an antiperiodic circle and 0A compactified on $S^1$ have $S$-conjugate $\Gamma_0(2)$ symmetries
\begin{align}
    \Gamma^0(2) = S \Gamma_0(2) S^{-1}.
\end{align}
In particular, the shift symmetry of $\tau$ is by two units in the 0A realization and by one unit in the IIA realization.
This suggests that these two realizations of 0B may not be non-perturbatively equivalent but mapped to one another under strong-weak duality.  In particular the direction in the M-theory torus with antiperiodic boundary condition is mapped to $S^1\vee S^1$ direction.  It is conceivable that the actual symmetry of 0B in 10d is enhanced to $SL(2,\bZ)$, but this is not required.  Nevertheless we find evidence below that indeed $SL(2,\bZ)$ is the duality symmetry that acts non-trivially on tachyon vevs as well and permutes the three infinity limits of $\Gamma_0(2)$, predicting that they are equivalent.

\subsection{Critical point of 0B}
Consider the locus with $T=\chi'=0$ and consider the potential $V$ as a function of the axio-dilaton $\tau$: $V(\tau,\overline \tau)$.  We ask if there is a critical point for this potential.  As was already pointed out in \cite{Ginsparg:1986wr} (see also \cite{Mohseni:2025tig,Chen:2025rkb,Lust:2022mhk,Grimm:2024fip,Font:1990gx,Cvetic:1991qm,Kokorelis:2000yt,Gonzalo:2018guu}), at a symmetry point of moduli space where all moduli pick up charge the potential is automatically critical, because $dV=0$ by charge conservation, even if the charge carried by the modulus is discrete. In other words $V$ is gauge neutral, but the derivative of it in all directions would be charged and thus vanish.  It turns out that $\Gamma_0(2)$ has an order 4 element with an action of order 2 on the moduli with a fixed point.  This element (which is conjugate to $S$ transformation of $SL(2,\bZ)$) is given by
\begin{align}
    g=(TS)^{-1} S (TS)=
\begin{pmatrix}
1 & -1 \\
2 & -1
\end{pmatrix},
\end{align}
with the fixed point
\begin{align}
    \tau=\frac{\tau -1}{2\tau-1}\Rightarrow \tau= \frac{1}{2}(i+1).
\end{align}
Note that the derivative of $V$ with respect to the complexified tachyon field $T+i\chi'$ will also vanish due to the $\bZ_2$ quantum symmetry it carries.

It is natural to ask what the sign of $V$ is at this critical point, as well as its stability. If this is the only critical point of $V$, we can determine that $V>0$ and $V''<0$.
The reason for this is that we know that at the two boundaries of $\tau$ moduli space we obtain weakly coupled 0B theory and as we have discussed the value of $V$ is positive.\footnote{For this argument we only need to know that at least on one boundary $V>0$ and we do not need to assume this is the case at all boundaries.} So this critical point, if unique, must be a global maximum of $V$, thus it leads to an unstable dS point. This idea has been generalized to other Scherk-Schwarz type toroidal compactifications of M-theory leading to isolated critical points \cite{Baykara:2026jzs}.

Note that the existence of this critical point follows from the well-established dualities of M-theory and even though it is compatible with our 0A proposal, its validity does not require our picture to hold.

\subsection{A non-supersymmetric CFT in 4d?}
In the context of Type IIB string theory, the S-duality of superstrings leads to the S-duality of $N$ coincident D3-branes, as D3-branes are invariant under it, leading to the prediction of Montonen-Olive duality from the perspective of string dualities.  An important ingredient in this is that there is no potential for the axio-dilaton in IIB, so the underlying spacetime is stable.
In the context of the 0B theory it is natural to ask whether a similar story holds.  In particular on the locus where the $\bZ_2$ quantum symmetry is unbroken we have a $\Gamma_0(2)$ duality symmetry.
So we ask whether there are D3-branes which respect these symmetries. Indeed, $D3^\pm$branes are invariant under $\Gamma_0(2)$ for the same reason they are in the context of usual IIB (as the $\Gamma_0(2)$ does not act on the 4-form fields $D^\pm$).  But to preserve the $\bZ_2$ quantum symmetry which exchanges $D3^+\leftrightarrow D3^-$, we would need equal number of coincident $N$ $D3^+$branes and $N$ $D3^-$branes.  So it is natural to expect that this brane system will inherit the $\Gamma_0(2)$ duality of 0B.  Moreover,
for the background to be stationary we need to be at a point where $V'=0$ and we have identified such a point: $\tau=\frac{i+1}{2}$.  So we conjecture that this non-supersymmetric theory at this strong coupling constant for gauge fields is conformal.  This theory consists of $U(N)\times U(N)$ gauge theory with six scalars in $(Adj,1)\oplus (1,Adj)$ and four bifundamental Weyl fermions in $(N,\overline N)\oplus (\overline N,N)$.  This is in principle possible and is compatible with showing the lack of conformal symmetry for this theory at weak coupling \cite{Klebanov:1998yya,Klebanov:1999ch}.    Presumably a suitable flow will take us to the critical point we have found as an IR fixed point.  This leads to an appealing picture:
Since the 0B theory enjoys the $\Gamma_0(2)$ duality symmetry, the system of branes we are considering will inherit this symmetry as well.  Moreover the RG flow vector field acting on $\tau$ should be a duality invariant object as is the statement about the flow towards the IR and this is a duality frame invariant concept. This would then imply that if we go to the critical point of the potential at $\tau_0=(i+1)/2$ the RG flow vector field must vanish as no direction in $\tau$ would be invariant under the $\bZ_2$ gauge symmetry acting on $\delta \tau$.
This strengthens the argument for the existence of this conformal fixed point.

If we take unequal rank for the $D3^\pm$ branes it is natural to expect a cascading behavior \cite{Klebanov_2000}.  In such a case one would expect the tachyon field to also flow as the background no longer respects the quantum symmetry.  Indeed this also leads us to a natural expectation of what the full duality group for 0B is, which we turn to next.

\subsection{Full duality group for 0B}
It is natural to ask what the full duality group for 0B is.  We have argued that when the tachyon field is set to zero it is given by a level two subgroup of $SL(2,\bZ)$. The natural extension of this when tachyon takes a general vev is to view it as splitting the $\tau$ to two parts $\tau_\pm$.  In other words we could view the tachyon field as a point $z$ on a torus with origin (so more precisely two points with the relative position denoted by $z$). Indeed this is something which is also the expected duality group when we have a cascading behavior \cite{Klebanov_1998}.  The level two subgroup arises when we identify the vanishing of the tachyon field with the choice of a non-trivial mid-point of the torus.  Thus we conjecture that the full duality group of 0B is that of a torus with a pair of points parameterized by $z$:\footnote{Note that the quantum symmetry corresponds to $z\rightarrow -z$ without changing torus moduli.  In particular $z\in T^2/\bZ_2$.}
$$(z,\tau)\rightarrow \left(\frac{z}{c\tau+d},\frac{a\tau +b}{c\tau +d}\right)$$
and 
$$T+i\chi'\propto \left(\frac{z}{\tau}-\frac{1}{2}\right),$$
where locally $z-\frac{\tau}{2}\approx\frac{1}{2}(\tau_+-\tau_-).$
If this conjecture is correct it would suggest that there is another interpretation of the F-theory picture:
It would be the zero area limit of M-theory compactification on a torus with two points separated by $z$ identified.  This would be a natural generalization of the $S^1\vee S^1$ picture which can be viewed as a circle with two points identified.
\subsection{Tachyon condensation for 0B}

Tachyon condensation picture which we geometrized in the M-theory lift of 0A has a parallel geometrization in the 0B theory:  going in the tachyonic direction would change $\tau_+$ relative to $\tau_-$.  It is natural to expect that this corresponds to $R_-/R\rightarrow 0$.  In this limit the extra torus in the F-theory picture disappears and we are back to the usual F-theory picture for the IIB theory.  Thus we see that our proposal naturally leads to Type IIB as we condense the tachyon.  In terms of the torus and the pair of points parameterized by $z$ (denoting the tachyon vev), this can be interpreted as moving the points together sending $z\rightarrow 0$.

\section{Concluding thoughts}\label{sec:conclusion}

In this paper we have proposed that Type 0A is given by an exotic compactification of M-theory on $S^1\vee S^1$.  In this map, the modes of weakly coupled 0A theory can be naturally geometrized. 
This comes at the cost of having fields live on different resolutions of this space. 
This can naturally fit with the picture that weak coupling is sub-Planckian M-theory, so the quantum aspects would be crucial to make sense of the space.  Moreover, we have found evidence that the potential $V$ grows as the sizes of the circles grow and this points to an obstruction to get to a macroscopic version of the geometry thus avoiding a contradiction with the lack of geometric meaning to compactification on large $S^1\vee S^1$.

We have noted that in this geometric picture the tachyon condensation is equivalent to shrinking one of the circles leading back to M-theory on $S^1$, i.e. Type IIA theory.  We have provided what we feel is strong evidence for this conjecture.  We have also used this picture to study 0B theory, and its geometric explanation of its moduli also predicting some critical points at strong couplings as well as argue that the fate of its tachyon condensation is to lead to IIB theory.

It would be interesting to flesh this out further.  The conditions we have imposed on the M-theory fields are rather exotic and we need further evidence for this picture. Our choices were motivated by the structure of D-branes in the 0A theory.  However, more needs to be checked in this context. In particular studying the structure and potential existence of bound states for $D0^+,D0^-,\overline{D0^+},\overline{D0^-}$branes would lead to a deeper understanding of the M-theory picture and how it resolves the $S^1\vee S^1$ singularity.  Also the end point of tachyon condensation presumably involves stronger quantum corrections to the classical picture.  That this has to be the case is clear because the fields which would require equal momenta on both circles would be naturally frozen (as the non-vanishing momentum of the shrinking circle would become very massive) but somehow these should now be liberated to have momenta only along the surviving circle.  The even momenta along each circle are there because they can be viewed as a combination of $(1,1)+(1,-1)=(2,0)$.  Again, the D-brane bound state dynamics may shed light on this question and needs to be further investigated.

Our proposal emboldens us to look for a geometric explanation of other non-supersymmetric string theories and the fate of their tachyon condensation.  It is natural to reconsider all non-supersymmetric string theories anew.  Indeed we find that this leads to new insights into other non-supersymmetric string vacua \cite{Baykara:2026vdc}.  It could also lead to new insights for non-supersymmetric domain walls predicted by the cobordism conjecture \cite{McNamara:2019rup}.  For example, consider the domain wall which interpolates between IIA and ${\rm IIA}'$.  Given what we saw for 0A it is natural to expect the profile of the domain wall as we interpolate from one side to the other involves the M-theory circle crossing itself at a point and reconnecting differently changing the orientation on the other half.

It is also natural to look for generalizations of our exotic compactifications.  Indeed there is a natural generalization one may expect, for which type 0A and type 0B can be viewed as the $SU(2)$ version.  This is motivated by the fact that in both cases we seem to have {\it two} copies of the universe.  If we viewed the universe as a D-brane, this would have naturally led us to expect an $SU(2)$ gauge symmetry.  Indeed for both 0A and 0B the tachyon field can be identified with the holonomy of a flat $SU(2)$ gauge field! In particular for 0A we found that the tachyon field can be identified with an interval which can be identified with the holonomy of Cartan subgroup $U(1)\subset SU(2)$ modulo the Weyl group action leading to $S^1/\bZ_2$.  Similarly the F-theory version leading to the 0B theory would be a flat $SU(2)$ holonomy on $T^2$ leading to moduli which is $T^2/\bZ_2$, which can be identified with the complexified tachyon field $z$ we have found for 0B.  This would suggest that perhaps an ADE version can be relevant for describing multiple copies of the universe and how they interact with one another.  It would be exciting to look for these and better understand the group theoretic structure of this observation.
\acknowledgments

We would like to thank A. Bedroya, G. Bossard, S. Chen, M. Delgado, T. Dumitrescu, L. Eberhardt, H. Parra de Freitas, K. Intriligator and R. Mahajan for valuable discussions. 

The work of ZKB and CV is supported in part by a grant from the Simons Foundation (602883,CV) and a gift from the DellaPietra Foundation. The work of E.~D. was supported in part by the IRP UCMN France-USA.  
\appendix

\section{Duality of freely acting orbifolds}\label{app:orbifold}
Consider a Type A (IIA or 0A) or Type B (IIB or 0B) theory with a symmetry $g$. Put it on a circle with a $g$ holonomy $S^1_g$. This is equivalent to taking a regular circle $S^1$ and then doing a freely-acting orbifold by
\begin{align}
    g\circ T_{\delta},
\end{align}
where $T_\delta$ implements the shift on $S^1$ by $\delta=1/N$ for $g^N=1$. The translation acts on the circle momentum-winding states as
\begin{align}
    T_\delta \ket{n,w} = e^{2\pi i n/N}\ket{n,w}.
\end{align}

Under T-duality, momentum and winding modes exchange. We distinguish the Type A and B states by subscripts
\begin{align}
\begin{split}
    \mathcal T: \ket{n_A,w_A} &\mapsto \ket{n_B,w_B},\\
     R_A &\mapsto R_B
\end{split}
\end{align}
with $n_B=w_A,w_B=n_A$ and $R_B=\frac{1}{R_A}$.

Viewed in the T-dual variables, the shift $T_\delta$ acts on the winding states as
\begin{align}
    e^{2\pi i n_A/N} \ket{n_A,w_A} = e^{2\pi i w_B/N} \ket{n_B,w_B}.
\end{align}
Defining the winding gauge transformation
\begin{align}
    \Omega_\delta \ket{n,w} \equiv e^{2\pi i w \delta}\ket{n,w},
\end{align}
we have the equivalence
\begin{align}
    T_\delta \ket{n_A,w_A} = \Omega_\delta \ket{n_B,w_B}.
\end{align}
In other words,
\begin{align}
    \mathcal T T_\delta \mathcal T^{-1} = \Omega_\delta.
\end{align}

Therefore, we have
\begin{align}\label{eq:duality}
 \frac{\text{Type A on }S^1}{g\circ T_\delta} \Bigg\vert_{R} = \frac{\text{Type B on } S^1}{g\circ \Omega_\delta}\Bigg\vert_{\frac{1}{R}}.
\end{align}

As $R\to0$, we see that the RHS of \eqref{eq:duality} decompactifies. Since $\Omega_\delta$ acts only on the winding modes, which become infinitely massive, only the $g$ action survives. Therefore
\begin{align}
    {\text{Type A on }S^1 \over g\circ T_\delta}\Bigg\vert_{R\to 0} = {\text{Type B}\over g}.
\end{align}
The other limit is
\begin{align}
    {\text{Type A on }S^1 \over g\circ T_\delta}\Bigg\vert_{R\to \infty} = \text{Type A}.
\end{align}
So the freely acting orbifold interpolates between Type A and $\frac{\text{Type B}}{g}$.

There is another freely acting orbifold that interpolates between those two theories. Let $Q$ be the quantum symmetry associated to orbifolding by $g$. Then
\begin{align}
    {{\text{Type B}\over g}\text{ on }S^1 \over Q\circ T_\delta}\Bigg\vert_{R\to 0} = \frac{{\text{Type A} \over g}}{Q} =\text{Type A}
\end{align}
and
\begin{align}
    {{\text{Type B}\over g}\text{ on }S^1 \over Q\circ T_\delta}\Bigg\vert_{R\to \infty} = {\text{Type B}\over g}.
\end{align}

These two interpolating models should be identified with $R_A\sim1/R_B$. In particular, we claim
\begin{align}
    {\text{Type A on }S^1 \over g\circ T_\delta}\Bigg\vert_{R} = {{\text{Type B}\over g}\text{ on }S^1 \over Q\circ T_\delta}\Bigg\vert_{\frac{N}{R}},
\end{align}
or using \eqref{eq:duality} equivalently
\begin{align}\label{eq:winding-shift-duality}
    {\text{Type B on }S^1\over g\circ \Omega_\delta} \Bigg\vert_{R \over N} = {{\text{Type B}\over g}\text{ on }S^1 \over Q\circ T_\delta}\Bigg\vert_{R}.
\end{align}

To prove \eqref{eq:winding-shift-duality}, let
\begin{align}
    Z^{\mathrm{int}}_{a,b} = \mathrm{Tr}_{g^a}( g^b q^{L_0} \bar q^{\bar L_0})
\end{align}
be the partial traces of the internal part, and
\begin{align}
    Z^{S^1,\Omega}_{a,b} = \sum_{n\in \mathbb Z+\frac{a}{N}}\sum_{w\in \mathbb Z} e^{2\pi i b w/N} q^{\frac 1 2 p_L^2}\bar q^{\frac 1 2 p_R^2}
\end{align}
be the circle part with $\Omega_\delta$ insertions and twists, where
\begin{align}
    p_{L,R} = \frac{1}{\sqrt{2}}\left(\frac n R \pm wR\right).
\end{align}

Then we have
\begin{align}
    Z\left[ {\text{Type B on }S^1\over g\circ \Omega_\delta}\right] = \frac1N \sum_{a,b} Z_{a,b}^{\mathrm{int}} Z_{a,b}^{S^1,\Omega}.
\end{align}

We first use the fact that quantum symmetry orbifolds are discrete Fourier transforms of the original partition function
\begin{align}
    Z_{k,\ell}^{\mathrm{int}/g/Q} = \frac1N\sum_{a,b} e^{\frac{2\pi i}{N}(a\ell -bk)}Z_{a,b}^{\mathrm{int}}.
\end{align}

Similarly, the circle part is an inverse Fourier transform as
\begin{align}\label{eq:circle-Z-DFT}
    Z^{S^1,\Omega}_{a,b} =  \frac1N \sum_{k,\ell}e^{\frac{2\pi i}{N}(a \ell-bk)} Z_{k,\ell}^{S^1,T}.
\end{align}
We now show \eqref{eq:circle-Z-DFT}. Expand the RHS 
\begin{align}
    \frac1N\sum_{k,\ell}e^{\frac{2\pi i}{N}(a \ell-bk)} Z_{k,\ell}^{S^1,T}=\frac1N\sum_{k,\ell}e^{\frac{2\pi i}{N}(a \ell-bk)} \sum_{n\in\mathbb Z}\sum_{w\in \mathbb Z+\frac{k}{N}} e^{2\pi i n\ell/N} q^{\frac 1 2 p_L^2}\bar q^{\frac 1 2 p_R^2}\\ \label{eq:S1-sum}
    =\frac1N \sum_{n\in\mathbb Z}\sum_\ell e^{\frac{2\pi i}{N} \ell (a+n)}\sum_{k}e^{-\frac{2\pi i}{N}bk} \sum_{w\in \mathbb Z+\frac{k}{N}} q^{\frac 1 2 p_L^2}\bar q^{\frac 1 2 p_R^2}.
\end{align}
Note that the term outside is nonzero only for $a+n \equiv 0 \pmod{N}$. Defining $n'=n/N$, this means only terms with $n'\in \mathbb Z - \frac{a}{N}$ survive in the sum. Likewise we define $w' = Nw$ as well, so the phase becomes $e^{-\frac{2\pi i}{N}bk}=e^{-2\pi ib w'/N}$. The effect of this redefinition is a scaling of the radius from $R$ of $Z^{S^1,T}$ to $R'=\frac{R}{N}$ of $Z^{S^1,\Omega}$, since left/right momenta in terms of new variables are
\begin{align}
   p_{L,R}= \frac{1}{\sqrt{2}}\left({n\over R} \pm wR\right) = \frac{1}{\sqrt{2}}\left({n' \over {R\over N}} \pm w' \frac{R}{N}\right).
\end{align}
Rearranging the sum and reversing summation variables, we get
\begin{align}
    \eqref{eq:S1-sum} = \sum_{n'\in\mathbb Z+ \frac{a}{N}}\sum_{w'\in \mathbb Z} e^{2\pi ib w'/N}  q^{\frac 1 2 p_L^2}\bar q^{\frac 1 2 p_R^2}= Z_{a,b}^{S^1,\Omega}
\end{align}
as claimed.

Now we prove \eqref{eq:winding-shift-duality}
\begin{align}
    Z\left[ {\text{Type B on }S^1\over g\circ \Omega_\delta}\right] &= \frac1N \sum_{a,b} Z_{a,b}^{\mathrm{int}} Z_{a,b}^{S^1,\Omega}\\
    &=\frac1N\sum_{a,b} Z_{a,b}^{\mathrm{int}} \frac1N\sum_{k,\ell} e^{\frac{2\pi i}{N}(a\ell-bk)} Z_{k,\ell}^{S^1,T}\\
    &= \frac1N\sum_{k,\ell}\left(\frac1N \sum_{a,b} e^{\frac{2\pi i}{N} (a\ell-bk)}Z_{a,b}^{\mathrm{int}}\right)Z_{k,\ell}^{S^1,T}\\
    &=\frac1N\sum_{k,\ell} Z_{k,\ell}^{\mathrm{int}/g/Q} Z_{k,\ell}^{S^1,T} = Z\left[{{\text{Type B}\over g}\text{ on }S^1 \over Q\circ T_\delta}\right].
\end{align}

\bibliographystyle{JHEP}
\bibliography{biblio.bib}

\end{document}